\documentclass[11pt,a4paper]{article}
\pdfoutput=1
\usepackage{enumitem} 
\usepackage{amsmath} 
\usepackage{amssymb} 
\usepackage{geometry}
\usepackage{geometry}

\usepackage{amsmath}

\usepackage{upgreek} 

\makeatletter\@addtoreset{chapter}{part}\makeatother%
\usepackage[dvipdfmx]{graphicx}
\usepackage[normal,oneline]{caption2}
\usepackage{url}
\usepackage[affil-it]{authblk}
\usepackage{textgreek}
\usepackage{xcolor}
\usepackage{bm}
\usepackage{appendix}

\usepackage{lineno}
\numberwithin{equation}{section}

\usepackage{upgreek} 

\begin{document}

\providecommand{\keywords}[1]
{
  \small	
  \textbf{\textit{Keywords---}} #1
}

\title{Mapping of the Transient Electric Field Created by the Charge\\
Spreading on the Resistive Strip of Micromegas\\
used in the ATLAS New Small Wheels}


\author[a]{T.~Alexopoulos}
\author[a]{S.~Maltezos\thanks{Corresponding author: S.~Maltezos, National Technical University of Athens, 15780 Athens, Greece,\\
e-mail: maltezos@central.ntua.gr}}
\author[a]{K.~Patrinos}
\author[b]{G.~Iakovidis}
\author[a]{G.~Koutelieris}

\affil[a]{National Technical University of Athens\\
Department of Physics, 157$\,$80 Athens, Greece}

\affil[b]{Brookhaven National Laboratory, Upton, NY 11973, USA}

\date{}
\maketitle

\pagenumbering{arabic}

\newcommand{\beq}{\begin{equation}}
\newcommand{\eeq}{\end{equation}}

\newcommand{\ben}{\begin{eqnarray}}
\newcommand{\een}{\end{eqnarray}}

\begin{abstract}
\emph{Abstract.}
Resistive strips Micromegas are employed in the ATLAS New Small Wheel project. They have been already installed and operate in the experimental cavern of the ATLAS experiment at CERN. This work attempts to describe the mechanism of the surface electric charge spread on a resistive strip of the Micromegas detector and the created transient electric field. Moreover, an approach utilising the Telegraph partial differential equation, used in transmission lines, is discussed assuming a 2-layer geometry approximation. Finally, the transient electric field formation leading to the strong suppression of the sparks in the amplification region is calculated by using the accurate 3-layer geometry for the particular parameters of the Micromegas configuration. The spatial mapping of this field versus time is presented for the time slice of interest. 

\noindent
\end{abstract} \hspace{10pt}

{\keywords{New Small Wheel, Resistive Micromegas, Charge spread, Spark suppression}}

\section{Introduction}
A major upgrade of the ATLAS~\cite{atlas} Muon Spectrometer at CERN in view of the HL-LHC is the New Small Wheels (NSW)~\cite{nswTDR}. It consists of two detector systems, resistive Micromegas and sTGC. It has already been installed in the ATLAS experiment. The Micromegas detectors~\cite{giomataris}, together with the sTGC ones~\cite{sTGC}, constitute the 16 sectors in each NSW.

The NSW Micromegas are of resistive type by which an adequate suppression of the spark rate is achieved. The mechanism of spark suppression is based on the development of an electric field in the opposite direction with respect to the constant electric field in the amplification region~\cite{nswMMres,agarwala}. This technique results in an instantaneous drop of the overall electric field during the spark formation, and like that, the gas gain is sufficiently reduced. Therefore, the resistive Micromegas detectors can be characterized as spark-insensitive~\cite{phd_iakovidis}. However, the highly ionizing particles, produced in the LHC collisions, still leads to an increasing probability for spark occurrence.

In the present work we study some important aspects of the resistive strip Micromegas functionality (for the resistivity chosen in the NSW application) mainly, at the time of electric charge deposition and its spread along the resistive strips. In particular, we study analytically and computationally the spread of the electric point charge (essentially a bunch of electrons) on the surface of the resistive strips. Moreover, the created transient electric field in the surrounding space is studied. By this way we perform a mapping of the electric field, focusing mainly on the $E_z$ component of the electric field which is collinear and opposite to the constant electric field due to the amplification of the nominal high voltage on the resistive strips. The results are used for further study of the mechanism of sparks, their onset, the spatial development, and the probability of occurrence.

The present work is organized as follows: In Section~\ref{s2} the electrical configuration of the NSW Micromegas is described
and the charge density mechanism is studied in the simpler case without including the Micromesh grid-shape electrode. In Section~\ref{s3} the charge density variation versus time and its spread along a resistive strip in large times is investigated. 
In Section~\ref{s4}, a mapping of the perpendicular component of the created transient electric field is given. The electric field, included the Micromesh is quantified by 2-layer geometry approximation as well as with the accurate 3-layer geometry, of Section \ref{s5}, describes the dependence of the electric field on the resistive layer thickness. Finally, the conclusions are presented.

\section{Structure and functionality of the NSW Micromegas}
\label{s2}
\subsection{Electrode configuration and strip patterns}
\label{s2-1}
The NSW Micromegas detectors consist of two main regions, the conversion/drift region and the amplification region separated with a grounded metallic grid called ``Micromesh''. Both regions are filled with a gas mixture, nominally being the Ar+$5\%$CO$_2$+$2\%$iC$_4$H$_{10}$ for the NSW application. The Micromesh is woven configuration of metallic wires of diameter $30\,\upmu$m which are $70\,\upmu$m apart with a pitch of $100\,\upmu$m.
The appropriate-constant voltages are applied to the cathode plane and the resistive strips creating the low electric field (drift region) and the high one (amplification region). The role of Micromesh is threefold: a) to separate the two regions, b) to be transparent for the incident electrons produced in the conversion/drift region and c) to collect the positive ions produced during the avalanche.

\begin{figure}[!ht]
\centering
\includegraphics[width=11.0cm]{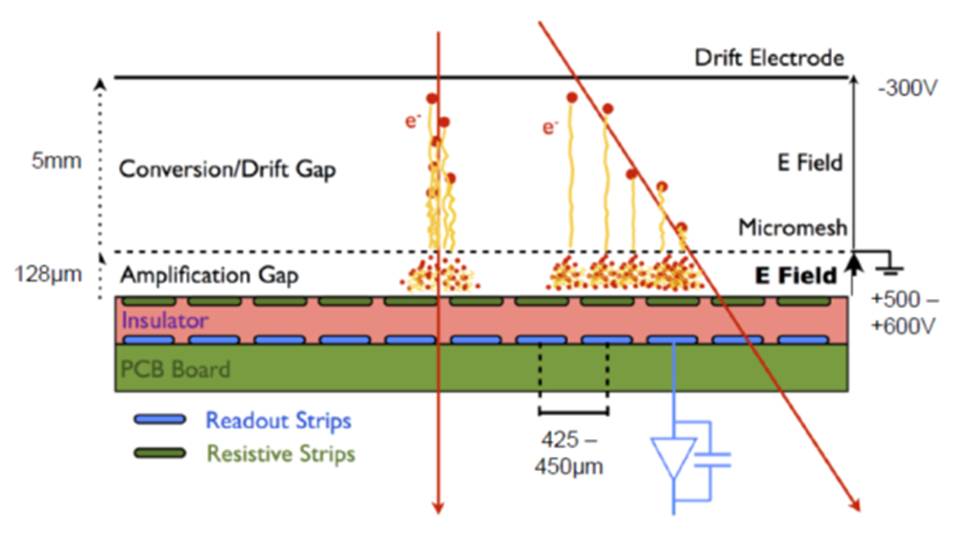}  
\caption{Micromegas perpendicular schematic layout and two representative particle tracks (perpendicular and inclined tracks).}
\label{mm_layout}
\end{figure}

\begin{figure}[!ht]
\centering
\includegraphics[width=8.0cm]{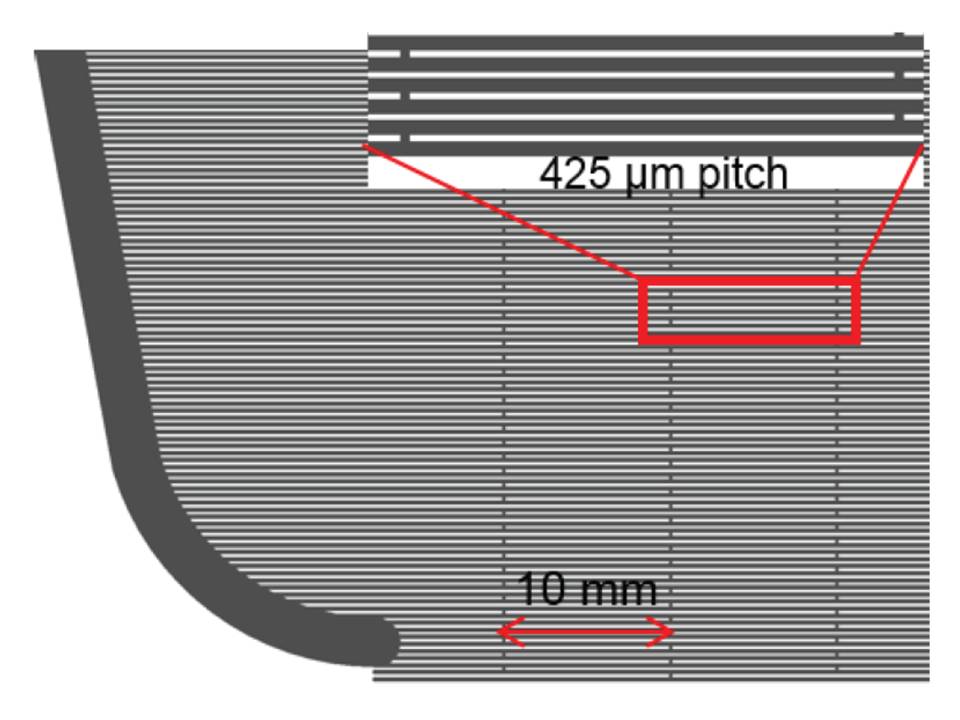}
\caption{Top view of the resistive pattern of NSW Micromegas. The periodic interconnections have not any impact to our analysis,
performed in small time slices, because of their large distances. However, they ensure the common potential in the entire pattern.} 
\label{resistive_pattern}
\end{figure}

For the charge collection and the signal formation, the detectors use a large number of resistive strips on top of a PCB layer and in a parallel configuration to the readout strips ($17\,\upmu$m thick, made of Cu)~\cite{nswMM}. Briefly, the overall mechanism is the following: the formation of the electron avalanches occur in the amplification region and the electric point charge is deposited almost instantaneously on a resistive strip, while signal is induced to the readout strips basically due to the motion of the charge. In the argon-based gas mixture used, about 50 electrons are liberated in the 5$\,$mm drift gap~\cite{phd_iakovidis}. The signal formed spreads over about 100-200\,ns due to the ion drift towards the cathode.
Between the resistive and readout strips there is an insulator of $64\,\upmu\mathrm{m}$ thickness. Each resistive strip is made of thin resistive film of thickness $d\approx 15\,\upmu$m, which presents a surface resistivity $R_\mathrm{s}$ measured in $\Omega/\square$ ($\Omega$ per square) which depends only on the material used and its thickness \cite{attie}. Namely, $R_\mathrm{s}=\rho d$, where $\rho$ is the electric resistivity of the material.
A representative side view schematic of the NSW Micromegas is shown in Fig. \ref{mm_layout} while in Fig. \ref{resistive_pattern}, the top view pattern of the resistive strips is also illustrated.

\begin{figure}[!ht]
\centering
\includegraphics[width=11.0cm]{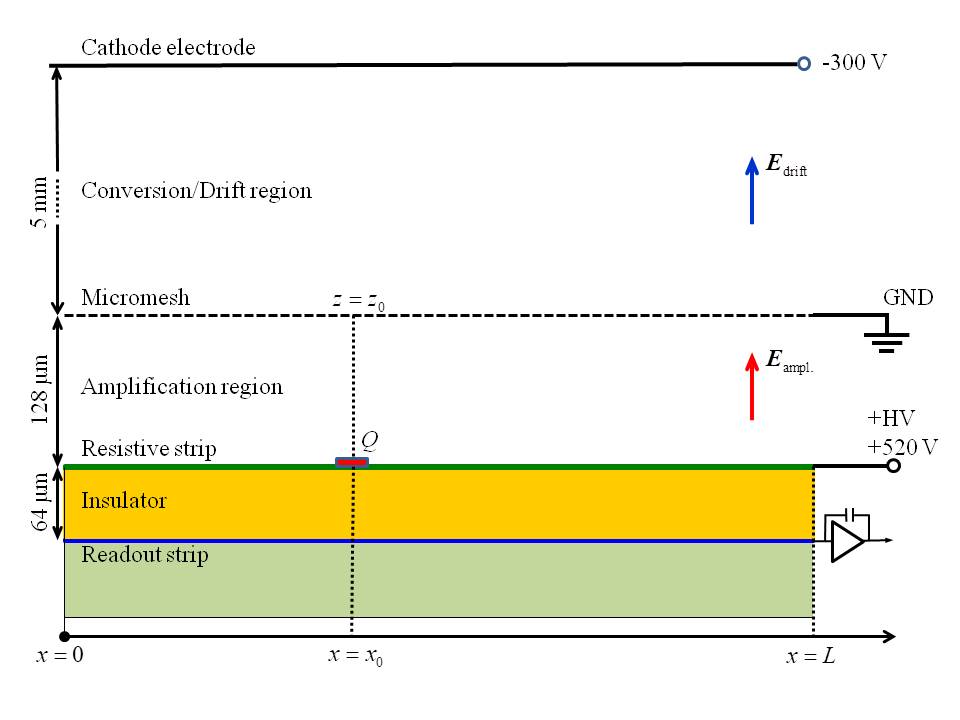} 
\caption{A schematic representation, not in scale, of the configuration of the drift and the amplification regions of a NSW Micromegas showing the resistive and the readout strips. The surface point charge is assumed to be deposited at an arbitrary position, $x=x_0,y=0$, along the resistive strip. The green region represents the PCB board consisting of a $500\,\upmu$m thick FR4 layer.}
\label{strip_conf}
\end{figure}

The geometric configuration of the above regions is shown schematically in Fig.~\ref{strip_conf} included more geometrical details. The Micromesh electrode is connected to the ground while the readout strips are connected to the ground through the electronics, namely through the charge sensitive amplifiers. Each resistive strip at one end ($x=0$) it is floating while at the other end ($x=L$) is connected together with all the others by using a conductor silver line which, in turn, is connected to the positive electrode (+HV) of the High Voltage power supply. In the first step of the analysis we concentrate to a small area of the resistive strip. At time $t=0$ we assume that an electric point-like charge cluster (or simply ``point charge'' used in the next) $Q$ is deposited instantaneously on a resistive strip at an arbitrary position $x=x_0$.  

\begin{figure}[!ht]
\centering
\includegraphics[width=10.0cm]{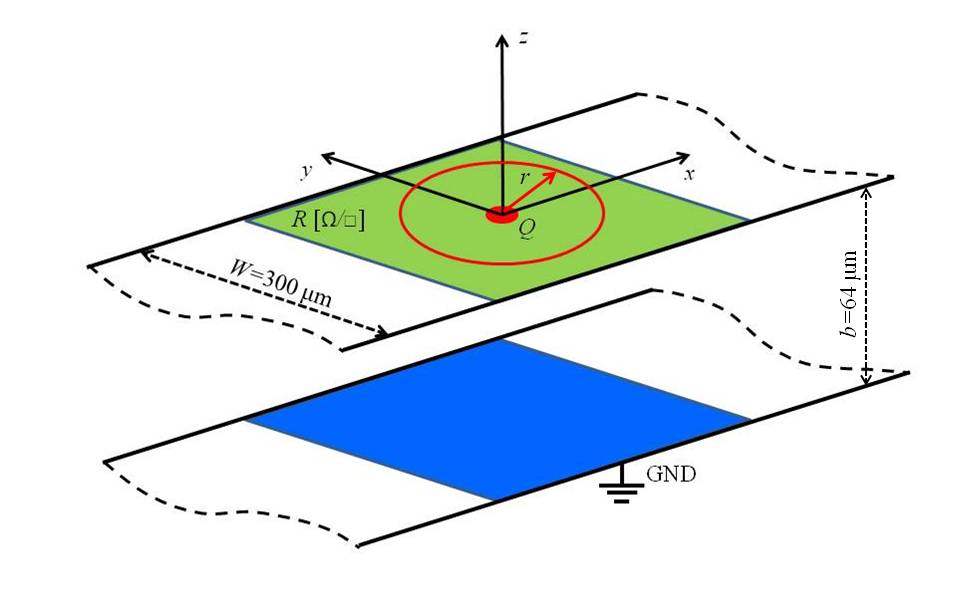} 
\caption{The part of the resistive strip along the $x$-axis where the surface point charge is deposited at $t=0$. The spread of the point charge is considered radially. The other parallel resistive strips are not shown for simplicity.}
\label{strip_radial}
\end{figure}

\subsection{Charge density spread mechanism}
\label{s2-2}
Let us assume that the resistive strip has a surface resistivity $R_\mathrm{s}$~\cite{maryniak,heaney,ochi}. Below the resistive strip, the corresponding readout strip extends in a parallel configuration and it is connected to the ground (assumed as a ground plane), as shown Fig.~\ref{strip_radial}. The analysis, initially, will be performed without the presence of a Micromesh in order to be able to investigate the charge spread mechanism in small and large times. The effect of the presence of the Micromesh to $E_\mathrm{z}$ component, is studied in the Section \ref{f5}, where we can verify that its effect concerns, a) in the magnitude depending on $z$ distance and b) in the decay time. Notably, the shape of the transient profile of $E_\mathrm{z}$ is similar in both cases. 

Eq. (6.6) in Section 6.1 of \cite{riegler} expresses the surface charge density in the approach of infinitely extended geometry of a resistive layer parallel to a grounded plane. The obtained non-zero coefficient $B_3$ contains the parameter $\nu$ (undetermined in the arbitrary geometry considered and having velocity units). In the case of infinitely extended geometry, the parameter $\nu$ is replaced by $1/2\varepsilon_0R$. In the NSW Micromegas, each PCB (left or right side) features 1024 resistive strips printed in parallel and inter-connected at their end $(x=L)$ to the +HV electrode. 

Before our analysis, it is worth to better clarify the overall ``picture'' of the electrical layout, included three electrode systems, a) Micromesh having a grid-shape which is grounded, b) the resistive strip system having a pattern with periodic interconnection among them and electrically connected together at their end by a silver line of +HV, as shown in Fig. \ref{resistive_pattern}, and c) the readout strip system considering grounded because each strip is connected to ground through the charge sensitive amplifier. 
These systems are extended over a large area of the order of square meters in each Micromegas module (PCB) including 1024 readout strips. Beside its segmented structure, in a macroscopic view the electric potential is well defined and can be used in the boundary conditions for solving the charge spread and transient electrical field problem. For better accuracy, we assume the charge deposit at the center of each resistive strip which operate independently. 

Considering a square area of these parallel strips with width $W$ we can, initially, use the result of \cite{riegler} based on infinitely extended geometry and calculated for resistive thickness infinitely small considering a 2-layer geometry. This configuration is similar to Micromegas without the Micromesh appropriately
adopted for the insulator used in our case. Regardless of the thickness $b$, according to \cite{riegler} (Fig. 24a and its legend) the grounded plane has a small effect on the charge density for small times. In our case of Micromegas strip configuration in $y$-direction we have:
$300\,\upmu\mathrm{m}/64\,\upmu\mathrm{m}\approx 4.7$, while in $x$-direction, with a length of the order of $1\,$m, is of the order of 15600.

According to the above assumption and based on the analysis given analytically in Section \ref{appendix} (Appendix), the expression for the charge density $q(r,t)$ becomes

\beq
q(r,t)\approx\frac{Q}{b^2\pi}\frac{1}{2}\int\limits_{0}^{+\infty }{\kappa {{J}_{0}}}\left( \kappa \frac{r}{b} \right){{e}^{-\kappa \left( 1-{{e}^{-2\kappa }} \right)t/T'}}\text{d}\kappa
\label{f1}
\eeq
\\
\noindent
where $J_0$ is the Bessel function, of first kind - zero order, $\kappa=bk$ is a dimensionless variable by definition based on the variable, $k$. Moreover, $T'$ is a ``characteristic time'', a constant which is defined by $T'=2b(\varepsilon_0 +\varepsilon)/2R_\mathrm{s}$, where $\varepsilon_0$ is the vacuum permittivity and $\varepsilon=\varepsilon_0 \varepsilon_r$ the permittivity of the insulator. We use this symbol $T'$, that is with prime, to avoid confusion with the symbol $T$ used in \cite{riegler}. 
By using $\varepsilon_r=3.55$ we obtain $T'=2.20$\,ns. In the Appendix the calculations for obtaining the above expression of $T'$ under some approximations are analytically presented.
The parameters $b$ and $R_\mathrm{s}$ are the thickness of the insulator and the surface resistivity, respectively. An approximation of first order by Taylor's expansion could be performed to the exponential function in the exponent 

\beq
{{e}^{-\kappa \left( 1-{{e}^{-2\kappa }} \right)t/T'}}\approx {{e}^{-2{{\kappa }^{2}}t/T'}}
\label{f12}
\eeq

\noindent The condition for performing the above approximation is explained in \cite{riegler} and is adequate for long times (namely for large values of $t/T'$). This is due, since the integral contributes only for small value of $\kappa$. In this case the integral of Eq.~\eqref{f1} becomes

\beq 
q(r,t)\approx \frac{Q}{b^2\pi}\frac{1}{2}\int\limits_{0}^{+\infty }{\kappa {{J}_{0}}}\left( \kappa \frac{r}{b} \right){{e}^{-2{{\kappa }^{2}}t/T'}}\text{d}\kappa
\eeq

\noindent This integral can be calculated analytically leading to the following Gaussian shape function

\beq
q(r,t)\approx \frac{Q}{b^2\pi}\frac{1}{8}{{\left( \frac{t}{T'} \right)}^{-1}}{{e}^{-\frac{{{r}^{2}}}{8{{b}^{2}}}{{\left( t/T' \right)}^{-1}}}}
\label{f13}
\eeq

\noindent Defining ${{t}_\mathrm{n}}\equiv t/T'$ as a normalized time, the charge density $q(r,t)$ can be described as follows  

\beq
q(r,{{t}_{n}})\approx \frac{Q}{b^2\pi}\frac{1}{8}{{e}^{-{{r}^{2}}/8{{b}^{2}}{{t}_{n}}}}
\eeq

\noindent The Gaussian analytic expression provides a significant advantage of fast computation time. This is necessary for the detailed mapping of the created electric field at small times, $t_\mathrm{n}<1$ or by using a practical value of time as a ``separation border'' $t_\mathrm{n}<t_\mathrm{s}$ for changing the expressions of charge density. By rewriting the exponential function as

\beq
{{e}^{-\kappa \left( 1-{{e}^{-2\kappa }} \right){{t}_{n}}}}={{e}^{-\kappa {{t}_{n}}+\kappa {{e}^{-2\kappa }}{{t}_{n}}}}\approx {{e}^{-\kappa {{t}_{n}}}}
\label{f2}
\eeq
\\
\noindent the second term contributes much less than the first one for $\kappa>>1$. For $\kappa$ around unity its contribution is still small being of the order of $1/e approx 0.37$. For $\kappa<<1$ the contribution becomes of the same order as that of the first term, but at the same time, their algebraic sum cancels out rapidly as soon as $\kappa$ tends to zero. Therefore, the approximation given in Eq.~\eqref{f2} is justified. Using this approximation, the charge density takes the following simplified form

\beq
q(r,{{t}_{n}})=\frac{Q}{b^2\pi}\frac{1}{2}\int\limits_{0}^{+\infty }{\kappa {{J}_{0}}}\left( \kappa \frac{r}{b} \right){{e}^{-\kappa {t_\mathrm{n}}}}\text{d}\kappa
\label{f3}
\eeq

\begin{figure}[!ht]
\centering
\begin{minipage}[b]{0.47\linewidth}
\centering
\includegraphics[scale=0.41]{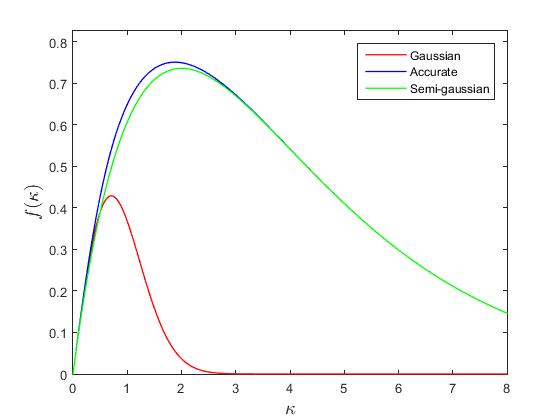}  
\caption{Graphical comparison of the three functions versus normalized time being used in the charge density integral assuming $t_\mathrm{n}=0.5$.}
\label{left_1}
\end{minipage}
\qquad
\begin{minipage}[b]{0.47\linewidth}
\centering
\includegraphics[scale=0.41]{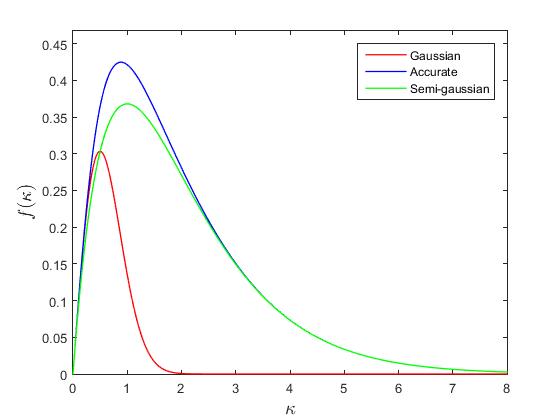}
\caption{Graphical comparison of the three functions as a function of distance being used in the charge density integral assuming $t_\mathrm{n}=1$.}
\label{right_1}
\end{minipage}
\end{figure}

\noindent In Fig.~\ref{left_1} and \ref{right_1} a comparison of the three functions, $f_a=\kappa {{e}^{-\kappa \left( 1-{{e}^{-2\kappa }} \right){t_\mathrm{n}}}}$, $f_g=\kappa {{e}^{-2{{\kappa }^{2}}t_\mathrm{n}}}$ and $f_{sg}=\kappa {{e}^{-\kappa {t_\mathrm{n}}}}$ is shown. The maximum deviation between the Semi-Gaussian and the Exact function is observed around $\kappa=1$. This is only $2\%$ higher for $t_\mathrm{n}=0.5$ while being $15\%$ for $t_\mathrm{n}=1$. The Gaussian function lags far behind the others in shape and scale.

\begin{figure}[!ht]
\centering
\begin{minipage}[b]{0.47\linewidth}
\centering
\includegraphics[scale=0.41]{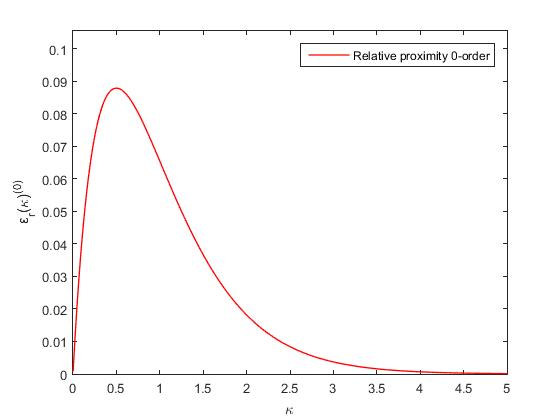}  
\caption{Plot of the relative proximity of zero order versus $\kappa$, for $t_\mathrm{n}=0.5$. The maximum approximation, corresponds to the maximum value at $\kappa=0.5$ and is $8.8\%$.}
\label{left_p1}
\end{minipage}
\qquad
\begin{minipage}[b]{0.47\linewidth}
\centering
\includegraphics[scale=0.41]{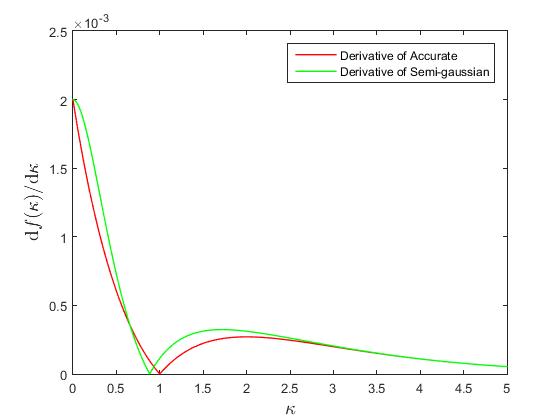}
\caption{Plot of the derivatives of the two functions, Semi-Gaussian and Accurate one versus $\kappa$ for $t_\mathrm{n}=0.5$. The maximum deviation is $20.6\%$ at $\kappa=0.2$.}
\label{right_p1}
\end{minipage}
\end{figure}

Mathematically, the two functions $f_a$ and $f_{sg}$, can be compared as follows. The absolute ``distance'' of two continuous functions, that is the ``proximity'', is  calculated with respect to the coordinates. This quantity must, at least, be calculated as zero order as well as of first order, defined as, ${{\varepsilon }^{(0)}}(\kappa )=\left| {{f}_{\text{a}}}(\kappa )-{{f}_{\text{sg}}}(\kappa ) \right|$ and ${{\varepsilon }^{(1)}}(\kappa )=\left| {{{{f}'}}_{\text{a}}}(\kappa )-{{{{f}'}}_{\text{sg}}}(\kappa ) \right|$, respectively. The zero order can be also written like: ${{\varepsilon }^{(0)}}(\kappa )=\kappa {{e}^{-\kappa {{t}_{n}}}}\left| \left( {{e}^{\kappa {{t}_{n}}{{e}^{-2\kappa }}}}-1 \right) \right|$. The plot of the relative proximity of zero order with respect to $f_a$ is shown in Fig.~\ref{left_p1}, while the plot of the derivative of the Semi-Gaussian function together with the derivative of the accurate function, is shown in Fig.~\ref{right_p1}.

\noindent Nevertheless, the above integral constitutes the following Laplace Transform identity where a Bessel function of first kind - zero order is included.

\beq
f(p)=\int\limits_{0}^{+\infty }{x{{J}_{0}}}\left( ax \right){{e}^{-px}}\text{d}x=\frac{p}{{{\left( {{a}^{2}}+{{p}^{2}} \right)}^{3/2}}}
\label{f4}
\eeq

\noindent Applying this expression to the function of Eq.~\eqref{f3} and setting $x\longrightarrow k$, $p\longrightarrow t_\mathrm{n}$ and $a\longrightarrow r/b$ we obtain the analytic expression of the integral

\beq
\int\limits_{0}^{+\infty }{\kappa {{J}_{0}}}\left( \kappa \frac{r}{b} \right){{e}^{-\kappa {{t}_{n}}}}\text{d}\kappa =\frac{{{t}_{\text{n}}}}{{{\left[ t_{\text{n}}^{2}+{{\left( r/b \right)}^{2}} \right]}^{3/2}}}
\label{f5}
\eeq

The expression of Eq. (\ref{f5}), substituting $t_\mathrm{n}=\nu t/b$, where $\nu=1/2\varepsilon_0 R_\mathrm{s}$, is of identical mathematical form with that of no presence of a grounded plane given in \cite{riegler} (Eq. (5.7)). Nevertheless, in our expression the normalized time is used, instead of $t$, and thus the parameter $b$ affects only the time scale. Moreover, the assumption of an infinitely extended resistive layer is not necessary to be imposed. Therefore, our approximation in the charge density for small times is approximately equivalent to a configuration not including a grounded plane.

\noindent Finally, we give the overall expression of the charge density in the entire range of space and time 
\beq
q(r,t_\mathrm{n})\approx\left\{
\begin{matrix}\displaystyle \frac{Q}{b^2\pi}\frac{1}{2}\frac{t_\mathrm{n}}{\left[t_\mathrm{n}^2+\left(r/b\right)^2\right]^{3/2}} & \textrm{for} & t_\mathrm{n} \le t_s \\ \; & \; & \; \\ \displaystyle \frac{Q}{b^2\pi t_\mathrm{n}}\frac{1}{8}\textrm{e}^{-r^2/8b^2t_\mathrm{n}} & \textrm{for} & t_\mathrm{n} >  t_s \\\end{matrix}\right.
\label{f6}
\eeq

\noindent where $t_\mathrm{s}=3.186$ has been specified by solving a transcendental equation for $r=0.01\,$cm as the time of intersection of the two approximated curves of the charge density. For higher accuracy we must set $t_\mathrm{s}=1$ and for the Gaussian approximation we must set $t_\mathrm{s}=10$. However, in the range of $1<t_\mathrm{s}<10$ there is a gap where only the exact integral can give a reliable result.
The derived analytic expression of the charge density is very useful for one more reason: even though it is an approximation according to the work in the~\cite{puoskari,lunaa}, the numerically calculated integrals contain the Bessel function first kind - zero order, as a rapidly oscillating function and a special method must be used.  

Based on the three functions investigated above, the integration was used according to the Eq.~\eqref{f1} for finding the charge density versus time and space utilizing the Matlab framework \cite{MATLAB}. In the calculations the following parameters for the NSW Micromegas were used: 

\begin{itemize}
\item{Deposited point-like surface charge at origin: $Q=16\,$pC}
\item{Thickness of the insulator between the resistive and readout strips: $b=64\,\upmu$m}
\item{Characteristic time of charge spread: $T'=2.20\,$ns}
\end{itemize}

\begin{figure}[!ht]
\centering
\begin{minipage}[b]{0.47\linewidth}
\centering
\includegraphics[scale=0.41]{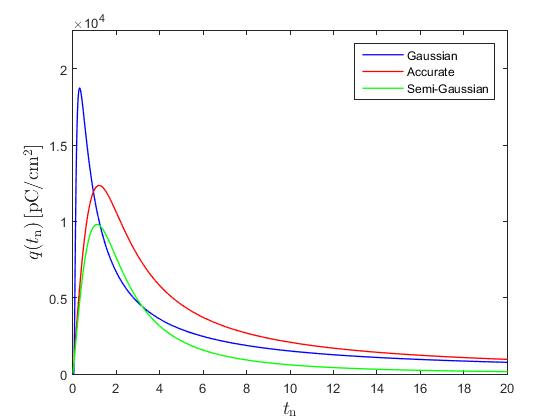}  
\caption[]{The charge density versus normalized time at the radial distance $r=0.01\,$cm.}
\label{left_d1}
\end{minipage}
\qquad
\begin{minipage}[b]{0.47\linewidth}
\centering
\includegraphics[scale=0.41]{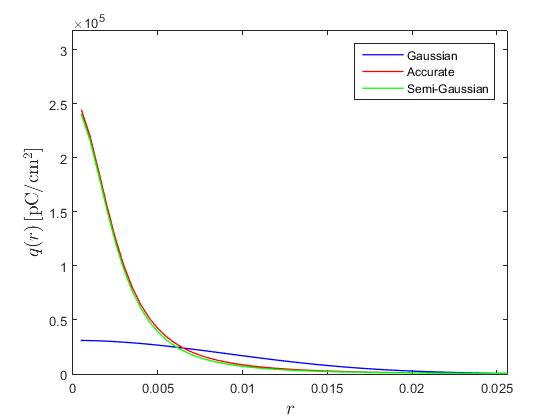}
\caption[]{The charge density versus radial distance at normalized time $t_\mathrm{n}=0.5$.}
\label{right_d1}
\end{minipage}
\end{figure}

\noindent In Fig.~\ref{left_d1} a plot of the charge density is shown for $r=0.01\,$cm versus normalized time $t_\mathrm{n}$. The Semi-Gaussian analytic integration (green curve) is close to the accurate integration (red curve) for small values of time. The Gaussian analytic result (blue curve) is a good approximation for large values of time at $t_\mathrm{n}>20$.
In Fig.~\ref{right_d1} a plot of the charge density versus radial distance for $t_\mathrm{n}=0.5$ is also shown. The very good approximation of the Semi-Gaussian analytic integration (green curve) is evident.  

\begin{figure}[!ht]
\centering
\begin{minipage}[b]{0.47\linewidth}
\centering
\includegraphics[scale=0.40]{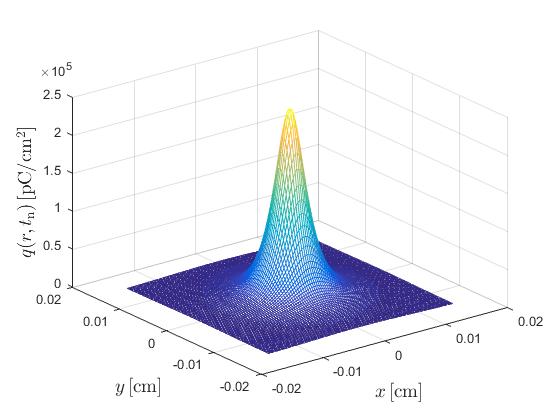}  
\caption{3D representation of the charge density spread along the resistive strip, at small times ($t_\mathrm{n}=0.5$), including two images. The Semi-Gaussian approximation for the integration has been used, while the actual time is $1.06\,$ns in the infinitely extended geometry approach approximation.}
\label{left_im}
\end{minipage}
\qquad
\begin{minipage}[b]{0.47\linewidth}
\centering
\includegraphics[scale=0.40]{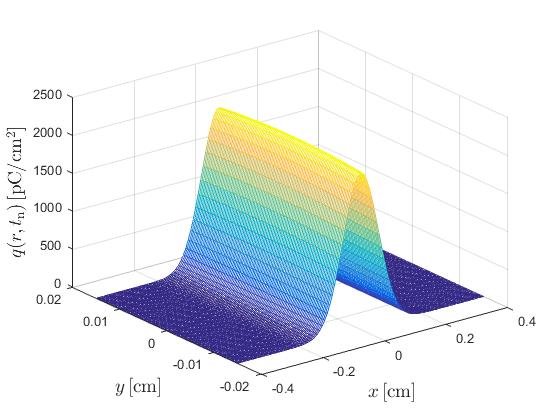}
\caption{3D representation of the charge density spread along the resistive strip, at large times ($t_\mathrm{n}=20$), including two images. The Gaussian approximation of solution has been used, while the actual time is $42.5\,$ns in the infinitely extended geometry approach approximation.}
\label{right_im}
\end{minipage}
\end{figure}

\noindent 
Since the resistive strip is much narrower in $y$-direction than in $x$-direction, the charge density spreads rapidly towards
the two ends of the strip for times $t_\mathrm{n}\approx 1$ (actual time $t\approx 2.2$\,ns). Therefore, the boundary conditions (Neumann type) at $y=-W/2$ and $y=W/2$ must be taken into account in an accurate 3D representation. By using the method of mirror images described analytically in~\cite{alexo_maltezos_2022}, two more virtual point charges at $t=0$ (of the same sign as the actual one) at the locations $(0,-W,0)$ and $(0,W,0)$ are included. At subsequent times the two mirror images become charge densities which in the general case of non even function must be also be converted by using reflection in the $y$-direction. By this method we can have a ``picture'' of the charge density spreading versus time. The results are presented in Fig.~\ref{left_im} and \ref{right_im} for small and large values of time, respectively. For the electric field mapping in the next Sections only the shape of the charge density at small times is used. 

An interesting case is when the point charge is deposited very close to the left end ($x=0$) or right end ($x=L$) of the resistive strip. In this configuration the solution of the charge spread versus space and time must take into account the corresponding boundary condition. 
If the point charge is deposited close to the left end it is spread mostly to the right while in the left direction a kind of reflection occurs due to the corresponding boundary condition (mathematically is realized by a virtual point charge - image).
If the point charge is deposited close to the right end it spreads mostly to the left while in the right direction, it undergoes a progressive loss due to the electrical connection to a conductor (+HV electrode) acting as a ``sink''. More details and the associated methodology are given in~\cite{alexo_maltezos_2022,francesco,crank,roisin}.

\section{Diffusion model of charge spread approximation in large times}
\label{s3}
\subsection{Transmission line approach}
\label{s3-1}
The above analysis of the charge density along the radial directions concerns small times and it's very useful for performing the
mapping of the transient electric field. Given though that the resistive strips in the case of NSW Micromegas are very long, of the order of one meter and very narrow ($W=300\,\upmu$m), it is reasonable to study a solution by using Cartesian coordinates, $x$ and $y$. 
In this approach the configuration of the resistive strip and the grounded readout strip is considered as a transmission line. The most general fundamental mathematical model to describe that is the one-dimensional hyperbolic second-order Telegraph equation~\cite{srivastava}

\beq
\frac{1}{LC}\frac{{{\partial }^{2}}u}{\partial {{x}^{2}}}=\frac{{{\partial }^{2}}u}{\partial {{t}^{2}}}+\frac{LG+RC}{LC}\frac{\partial u}{\partial t}+\frac{RG}{LC}u
\label{te}
\eeq
\\
\noindent where $u$ is the voltage across a transmission line at any position and any time. $R$ denotes the resistance of the cable, $C$ the capacitance to the ground, $L$ the inductance of the cable and $G$ the conductance to the ground, all considered per unit length.  
For using the charge instead of voltage in the above equation, we must use the relation $q=Cu$. It is a good approximation when the gradient of the charge density is small over the transverse dimension of the transmission line. Even the desired relationship is more complicated, the capacitance per unit length along the transmission line could be measured experimentally and used as an effective one in the above basic formula. 
Furthermore, it can be assumed that $G\approx0$ and thus Eq.~\eqref{te} is reduced to

\beq
\frac{1}{RC}\frac{{{\partial }^{2}}q}{\partial {{x}^{2}}}=\frac{L}{R}\frac{{{\partial }^{2}}q}{\partial {{t}^{2}}}+\frac{\partial q}{\partial t}
\eeq

\noindent Assuming that the term $(L/R){{\partial }^{2}}q/\partial {{t}^{2}}$ is very small compared to the contribution of the other terms, mainly due to small inductance $L$ and high resistance $R$ as well as in the case of slow temporal variations of $q$, this equation takes the form of diffusion-like equation including the relevant diffusion constant given by $D=1/RC$, in units of area versus time \cite{dixit1,dixit2}. The resistance $R$ relates to a deposited layer of charge on the resistive strip 

\beq
\frac{\partial q(x,t)}{\partial t}=D\frac{{{\partial }^{2}}q(x,t)}{\partial {{x}^{2}}}
\eeq

A reasonable question arises: how we could relate this diffusion constant with that defined in the case of the radial model analyzed in the previous section? The answer to this question can be given by referring to Section \ref{s2} where a surface resistivity has been used to calculate the characteristic time $T'$. Based on the configuration of Fig.~\ref{strip_radial} the surface capacitance is given by $C_\mathrm{s}=\varepsilon/b$ in units of capacitance per unit area, where the index ``s'' denotes the arbitrary surface. Therefore, the corresponding diffusion constant should be

\beq
D_\mathrm{s}=\frac{1}{R_\mathrm{s}C_\mathrm{s}}=\frac{b}{{{\varepsilon }}{{R}_{\text{s}}}}
\eeq      
\\
\noindent where $\varepsilon=\varepsilon_r\varepsilon_0$ is the permittivity of the insulator. Furthermore, the relationship between $R$ and $R_\mathrm{s}$ is, $R_\mathrm{s}=WR$, where $W$ is the width of a long strip. Therefore, we have

\beq
D=\frac{1}{RC}=\frac{W}{{{R}_{\text{s}}}C}
\eeq      

\noindent These diffusion constants, concerns a homogeneous and isotropic material (the resistive layer in our case) and therefore we
equalize them to obtain an expression of the capacitance of the strip per unit length, as follows

\beq
{{D}_{\text{s}}}=D\Rightarrow \frac{b}{\varepsilon {{R}_{\text{s}}}}=\frac{W}{{{R}_{\text{s}}}C}
\eeq      
\\
\noindent leading to 

\beq
C=\frac{\varepsilon }{b}=W{{C}_{\text{s}}}
\eeq      

This approach for studying the charge density versus space and time can be realized under the assumptions
including the boundary conditions in $x$ and $y$ direction. 
Putting all together and referring to a typical NSW Micromegas configuration ($b=128\,\upmu$m, $\varepsilon_r=3.55$, $R=8\,$M$\mathrm{\Omega}$/cm), the above parameters take the following values:

\begin{enumerate}[label=\alph*)]

\item Approach assuming \textit{charge spreading along a surface}
\begin{enumerate}[label=\arabic*)]
\item{$R_\text{s}=2.4 \times 10^5\,\mathrm{\Omega}/\square$=0.24$\,\mathrm{M}\mathrm{\Omega}/\square$}
\item{$C_\text{s}=4.91\times 10^{-7}\,$F/m$^2$=49.1$\,$pF/cm$^2$}
\item{$R_\text{s}C_\text{s}=0.118\,$s/m$^2$}
\item{$D_\text{s}=1/R_\text{s}C_\text{s}=8.48\,$m/s$^2=0.0848\,$cm$^2$/$\upmu$s$\approx 1/11.8\,$cm$^2$/$\upmu$s}
\end{enumerate}

\item Approach assuming \textit{charge spreading along transmission line}
\begin{enumerate}[label=\arabic*)]
\item{$R=8\times 10^{6}\,\mathrm{\Omega}$/cm=8$\,$M$\mathrm{\Omega}$/cm}
\item{$C=1.47\times 10^{-10}\,$F/m=$147\,$pF/m}
\item{$RC=R_\text{s}C_\text{s}$}
\item{$D=D_\text{s}$}
\end{enumerate}

\end{enumerate}

\begin{figure}[!ht]
\centering
\includegraphics[width=13.0cm]{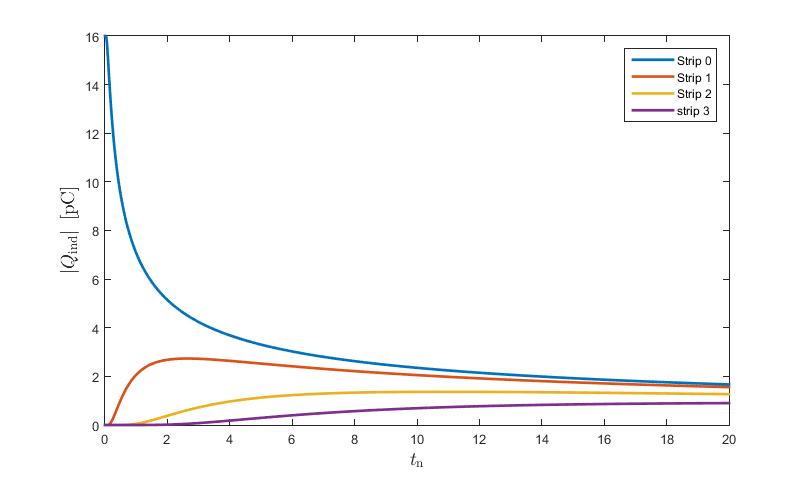} 
\caption{Plot of the absolute value of the induced charge along four readout strips versus time based on the strip pitch of $425\,\upmu$m for strip 1, $850\,\upmu$m for strip 2 and $1275\,\upmu$m for strip 3 with respect to the central strip. The charge maximum fractions with respect to the central strip are, $17.1\%$, $8.54\%$ and $5.63\%$, respectively. The half-time for the central strip corresponds to $t_\mathrm{n}=0.75$ which represents an actual time of $1.65\,$ns.}
\label{q_ind}
\end{figure}

We must notice that the capacitance per unit length, $C$, for the NSW Micromegas strip configuration was measured at $C\approx100\pm 20\,$pF/m. The difference might be attributed to the data of $\varepsilon_r$ used in the calculation of $C_\text{s}$ combined with the
measurement uncertainty of $C$.
According to the obtained charge density spread versus time, we can assume that the diffusion-like spread along $y$ direction is done in very small times of the order of ns leading to a uniform charge density almost instantaneously. Thus, the charge spread takes place only along $x$.  

\subsection{Induced charge cross-talk among readout strips versus time}
\label{s3-2}
Another significant study is the calculation of the induced charge from a resistive strip, let us call it central resistive strip (strip 0), to the corresponding readout strip being parallel and in the same pattern as well as to the three nearby strips (strip 1, strip 2 and strip 3). The induced charge is distributed at the readout strips located on both sides, left and right with respect to strip 0. Due to symmetry, the calculation can be done only on one side. According to the NSW Micromegas configuration, the pitch of the resistive and readout strips is $425\,\upmu$m and the gap is $\left( 425-300\right)\,\upmu\mathrm{m}=125\,\upmu$m and therefore only a fraction of the total charge along a resistive strip is induced to the total array of the nearby readout strips. 
The calculations were based on the methodology given in \cite{riegler} (Section 6.1) working in orthogonal coordinates and adapted for the configuration of the NSW Micromegas. The induced charge versus time is deduced applying the Gauss's law on the electric field $E_\mathrm{z}$ and integrating in $x$ and $y$ coordinates. This is done taking into account the lower and upper limits for the corresponding locations of the strip under study. In order to have an analytic expression, the approximated expression of the charge spread model presented in Section.~\ref{s2-2} has been used. This approximation holds for large times as well as for intermediate small values of time, in units of normalized time ($t_\mathrm{n}$) while a prompt use of the Gaussian approximation of the charge spread according to Eq.~\eqref{f13} fails for small values of time. This analytic approximate expression is 

\beq
\left| {{Q}_{\text{ind}}}({{t}_{\text{n}}}) \right|\approx \Theta ({{t}_{\text{n}}})\frac{Q}{2}\left[ \text{erf}\left( \frac{2{{y}_{\text{P}}}+W}{4b\sqrt{2{{t}_{\text{n}}}}} \right)-\text{erf}\left( \frac{2{{y}_{\text{P}}}-W}{4b\sqrt{2{{t}_{\text{n}}}}} \right) \right]
\eeq
\\
\noindent where $\Theta ({{t}_{\text{n}}})$ is the unit step function (Heaviside Function) representing the deposited point charge $Q$ at time $t=0$. By using the values of the parameters, $b$, $W$ and $y_\mathrm{p}$ for the location of each readout strip, we obtained the curve families of the induced charge (their absolute values) versus time presented in Fig.~\ref{q_ind}.
In this plot, one can observe that at time $t=0$ the summation of the induced charge is indeed equal to the initial deposited ($Q$), while at later time, it is smaller due to the gaps between the parallel readout strips. The expected fraction of the initial point charge $Q$ distributed among the readout strips versus time can be calculated (for both sides) as follows: 
if $P$ is the pitch and $n$ is the total number of readout strips, the expression providing this fraction is
$f_\mathrm{iq}=nW/\left[(n-1)P+W\right]$. Assuming $n>>1$ we obtain $f_\mathrm{iq}\approx W/P=0.7059$.

For one side the expected induced charge should be $0.7059\times 16/2=5.647\,$pC. This is the expected total induced charge for very large values of time. Based on the plot, we obtain slightly less values due to the small number of readout strips that were included in the calculation. We must notice here that for values of time about $t_\mathrm{n}=20$, the plot of Fig. \ref{q_ind} should not be realistic because the method of images, given in \cite{alexo_maltezos_2022}, must be applied for including the effect of the boundary conditions at the two sides in $y$-direction.

\section{Mapping of the electric field versus time}
\label{s4}
The mapping of the created electric field is the main goal of this work. Below we describe the methodology of
the calculations minimizing the computing power and time. 
According to the obtained results, the charge density is given by expressions which have circular symmetry. This allows us, at first, to calculate analytically the perpendicular electric field component, $E_\mathrm{z}$, at a distance of $128\,\upmu$m on the $z$-axis and at an observation point, P$_0$, as well as in several points below it, P$_i$ ($i=1,...N$). In the next stage, the electric field mapping around the point P$(0,0,z$) is performed by using a method of calculating the electric field promptly by Coulomb's Law. In both stages we use the approximated expressions of the charge density, for small and large times, according to the following methodology.

For the calculation of the component $E_\mathrm{z}(0,0,z)$, we consider small and large values of time working as follows: The infinitesimal electric charge at P$(0,0,0$),  is $2\pi q(r,t)\mathrm{d}r$ which the origin of the radial charge density. The component $E_\mathrm{z}(0,0,z,t)$ is calculated by integrating along the charge density area. Taking into account that $r^2=r'^2+z^2$, where $r'$ represents the radial distance on the resistive strip, we have 

\beq
{{E}_{\text{z}}}(0,0,z,t)=\frac{1}{4\pi {{\varepsilon }_{0}}}\int\limits_{0}^{+\infty }{\frac{\cos \varphi }{r{{'}^{2}}+{{z}^{2}}}q(r',t)}2\pi r'\text{d}r'=\frac{1}{2{{\varepsilon }_{0}}}\int\limits_{0}^{+\infty }{\frac{z}{{{\left( r{{'}^{2}}+{{z}^{2}} \right)}^{3/2}}}q(r',t)}r'\text{d}r'
\label{f10}
\eeq
\\
\noindent where $\varphi$ is the angle between the position vector $r$ and the $z$-axis which is the same for all elementary charges lying on each ring of width $\mathrm{d}r'$.
Substituting the expression of the charge density given by the Eq.~\eqref{f5} we obtain

\beq
{{E}_{\text{z}}}(0,0,z,{{t}_{\text{n}}})=\frac{Q{{t}_{\text{n}}}}{ 4{{b}^{2}}\pi {{\varepsilon }_{0}}}\int\limits_{0}^{+\infty }{\frac{r'z}{{{\left( r{{'}^{2}}+{{z}^{2}} \right)}^{3/2}}{{\left[ t_{\text{n}}^{2}+{{\left( r'/b \right)}^{2}} \right]}^{3/2}}}\text{d}r'}
\label{f11}
\eeq

\begin{figure}[!ht]
\centering
\includegraphics[width=8.5cm]{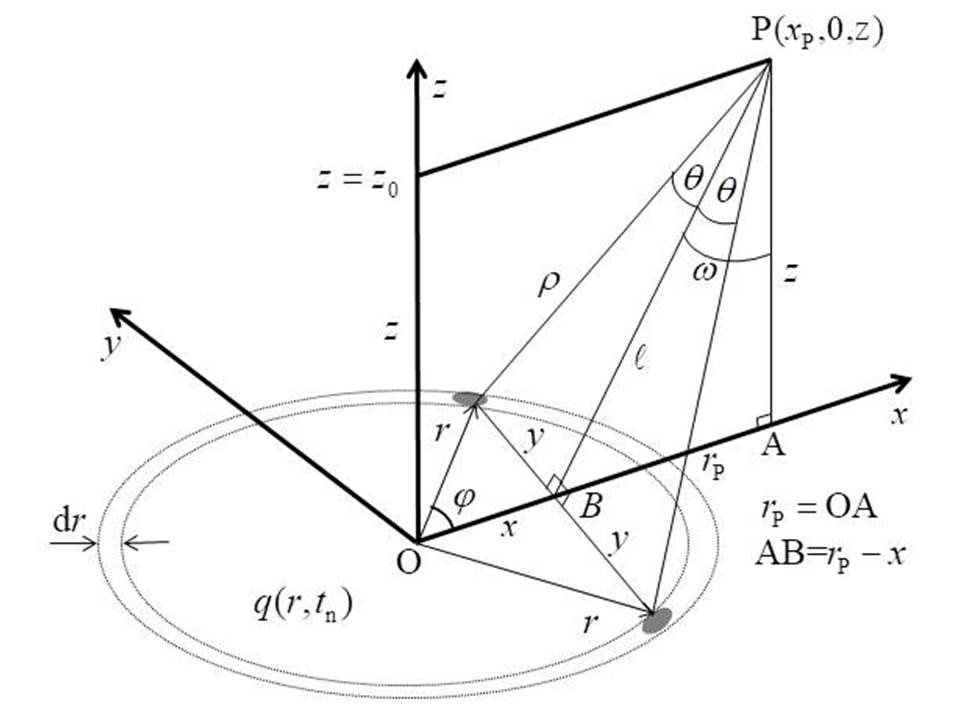} 
\caption[]{A schematic showing the geometrical configuration regarding the electric field created by an arbitrary radial charge density.}
\label{ez_geom}
\end{figure}

\begin{figure}[!ht]
\centering
\begin{minipage}[b]{0.47\linewidth}
\centering
\includegraphics[scale=0.40]{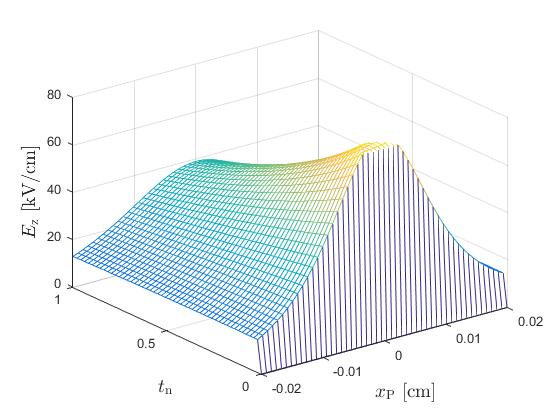}  
\caption[]{Space-time variation in 3D plot of $E_\mathrm{z}$ calculated at the observation point P.}
\label{left_ezm}
\end{minipage}
\qquad
\begin{minipage}[b]{0.47\linewidth}
\centering
\includegraphics[scale=0.40]{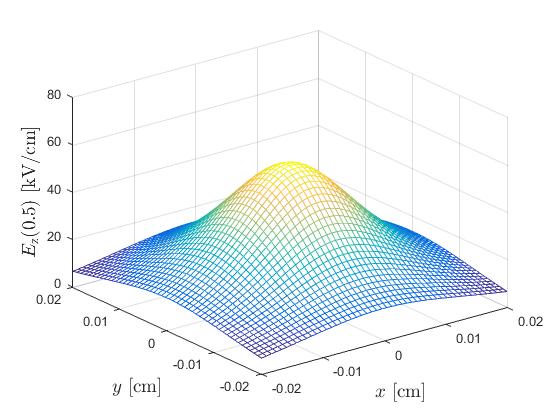}
\caption[]{Drop of $E_\mathrm{z}$ in 3D plot along nearby radial directions for small times ($t_\mathrm{n}=0.5$).}
\label{right_ezm}
\end{minipage}
\end{figure}

For small values of time the above integral is calculated setting $t_\mathrm{n}=0.5$ while for large values of time the Gaussian approximation is used setting $t_\mathrm{n}=20$. The integral takes the following forms

\beq
{{E}_{\text{z}}}(0,0,z,0.5)=\frac{Q}{8{{b}^{2}}\pi {{\varepsilon }_{0}}}\int\limits_{0}^{+\infty }{\frac{r'z}{{{\left( r{{'}^{2}}+{{z}^{2}} \right)}^{3/2}}{{\left[ 0.25+{{\left( r'/b \right)}^{2}} \right]}^{3/2}}}\text{d}r'}
\label{f14}
\eeq

\noindent For large times it is obtained accordingly

\beq
{{E}_{\text{z}}}(0,0,z,20)=\frac{Q}{320{{b}^{2}}\pi {{\varepsilon }_{0}}}\int\limits_{0}^{+\infty }{{{e}^{-r{{'}^{2}}/160{{b}^{2}}}}\frac{r'z}{{{\left( r{{'}^{2}}+{{z}^{2}} \right)}^{3/2}}}}\text{d}r'
\label{f15}
\eeq

For performing the mapping of the created electric field, $E(x,y,z,t_\mathrm{n})$, the following methodology has been implemented: due to the radial symmetry of the charge distribution, the positions of calculation can have an arbitrary direction around the observation point, let P$(r_\mathrm{P},0,z_0)$, where $z_0=128\,\upmu$m as shown in Fig. \ref{ez_geom}. The electric field $E_\mathrm{z}$ is calculated as a double projection, as follows: assuming two infinitesimal electric charges of the charge density, the electric field has the direction of $\rho$-line. The successive projection, to the $\ell$-line (angle $\theta$) and afterwards to $z$-axis ($\omega$), provide the component $E_\mathrm{z}$. For the infinitesimal quantities we can write ${{\text{d}}^{2}}{{E}_{\text{z}}}(0,0,z,{{t}_{\text{n}}})={{\text{d}}^{2}}\left[ E(0,0,z,{{t}_{\text{n}}})\cos \theta \cos \omega  \right]$. By integrating along the entire charge distribution we get

\beq
{{E}_{\text{z}}}(0,0,z,{{t}_{\text{n}}})=\frac{1}{4\pi {{\varepsilon }_{0}}}\iint{\text{d}S\frac{q(r,{{t}_{\text{n}}})}{{{\rho }^{2}}}\cos \theta \cos \omega }
\eeq
\\
\noindent where $S$ is the entire area of the spreading charge density versus time, theoretically infinite. In this approach the integration can be calculated (numerically) along the two corresponding semi-circles, and therefore, the result should be twice the double integral of only one semi-circle. Based on the corresponding right triangle we can write
 
\beq
{{\rho }^{2}}={{r}^{2}}+r_{P}^{2}-2{{r}_{P}}r\cos \varphi +{{z}^{2}}
\label{ge1}
\eeq

\begin{figure}[!ht]
\centering
\includegraphics[width=13.0cm]{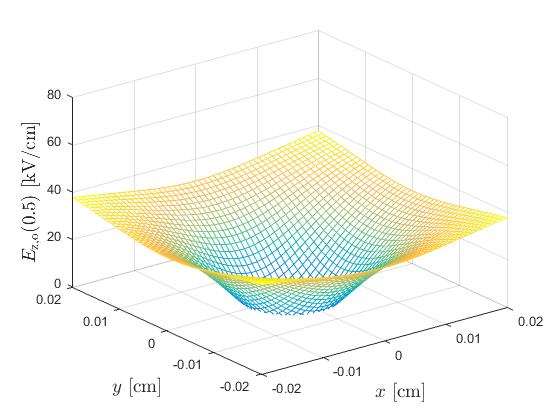} 
\caption{The crater-shape in the 3D plot of the overall transient electric field at $z=128\,\upmu\mathrm{m}$ and nearby radial distances above the charge distribution center for $t_\mathrm{n}=0.5$ or actual time $1.1$\,ns.}
\label{ezo}
\end{figure}

\noindent Based on this and the right triangle including the angle $\theta$ as well as the $x=r\sin{\varphi}$ and since $\sin \theta =y/\rho=r\sin \varphi/\rho$ we can also write 

\beq
\cos \theta =\sqrt{1-\frac{{{r}^{2}}\sin \varphi }{{{\rho }^{2}}}}=\sqrt{1-\frac{{{r}^{2}}\sin \varphi }{{{r}^{2}}+r_\mathrm{{P}}^{2}-2{{r}_{\mathrm{P}}}r\cos \varphi +{{z}^{2}}}}
\label{ge2}
\eeq

\noindent Additionally, $(AB)^{2}=r_{P}^{2}+{{x}^{2}}-2x{{r}_{\text{P}}}$ and using $x=r \cos{\varphi}$ we obtain

\beq
(AB)^{2}=r_{\text{\text{P}}}^{2}+{{r}^{2}}{{\cos }^{2}}\varphi -2x{{r}_{\text{P}}}
\eeq

\noindent and from the right triangle including the angle $\omega$ we obtain

\beq
\cos \omega =\frac{z}{{{(AB)}^{2}}+{{z}^{2}}}=\frac{z}{r_{\text{P}}^{2}+{{r}^{2}}{{\cos }^{2}}\varphi -2x{{r}_{\text{P}}}+{{z}^{2}}}=\frac{z}{r_{\text{P}}^{2}+{{r}^{2}}(1-{{\sin }^{2}}\varphi )-2r\cos \varphi {{r}_{\text{P}}}+{{z}^{2}}}
\label{ge3}
\eeq

Combining the geometrical relationships of Eq.~\eqref{ge1}, \eqref{ge2} and \eqref{ge3} the following expression of the integral is derived where the integration is performed along $r$ and $\varphi$.

\beq
{{E}_{\text{z}}}(0,0,z,{{t}_{\text{n}}})=\frac{1}{2\pi {{\varepsilon }_{0}}}\int\limits_{0}^{\pi }{\text{d}\varphi }\int\limits_{0}^{+\infty }{\text{d}r\frac{rq(r,{{t}_{\text{n}}})}{{{\rho }^{2}}}\cos \theta \cos \omega }
\eeq

\noindent where
\begin{eqnarray} q(r,t_\mathrm{n}) = \frac{Qt_\mathrm{n}}{2\pi b^2} \left[{t_\mathrm{n}}^2 + \left(\frac{r}{b}\right)^2\right]^{-3/2}\ \textrm{for}\ t_\mathrm{n} \le t_s,\ \textrm{and}\ {{\rho }}=\sqrt{{{r}^{2}}+r_{\text{P}}^{2}-2{{r}_{\text{P}}}r\cos \varphi +{{z}^{2}}} \nonumber
\end{eqnarray}

In Fig. \ref{left_ezm} and \ref{right_ezm} the electric field $E_\mathrm{z}$ is given in 3D plots versus space and time as well as in radial distances at a given normalized time, respectively. 
The direction of $E_\mathrm{z}$ is opposite to the constant electric field applied in the amplification region, which is $E_\mathrm{ampl}=44.5$\,kV/cm. Therefore, the overall transient electric field should be $E_\mathrm{z,o}=E_\mathrm{ampl}-E_\mathrm{z}$, which is shown in Fig. \ref{ezo} as a function of radial distance.

\section{The electric field including the Micromesh}
\label{s5}
\subsection{2-layer geometry approximation}
\label{s5-1}
\begin{figure}[!ht]
\centering
\includegraphics[width=16.0cm]{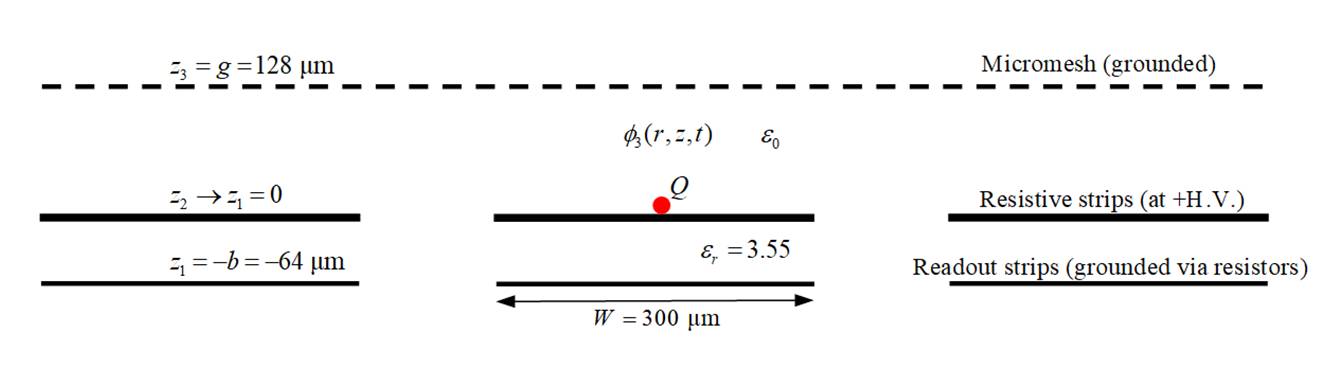} 
\caption{A side view scheme of the configuration of three indicative and neighborhood Micromegas electrodes and strips. The geometrical dimensions of the strips and distances are in relative scales. Our analysis focuses to the central region of any strip.}
\label{figapp1}
\end{figure}

The Micromesh is a metallic grid, which is expected to affect the electrical field at all times during the charge spread. At time $t=0$ the surface point charge is described mathematically by a Dirac - $\delta$ function $Q\delta(r)/2\pi r$ and thus its integral along $r$ is equal to the deposited total charge $Q$. In \cite{riegler}, the problem of finding the potential in a geometry consisting of two layers of different permittivity inside two grounded planes is studied, solving the Laplace equation by using cylindrical coordinates. Taking into account the boundary conditions, $\varPhi(r,z_3)=0$ (at grounded Micromesh plane), $\varPhi(r,-b)=0$ (at grounded readout strip plane) and that of the deposited charge, according to the configuration shown in Fig. \ref{figapp1}, the potential is written

\beq
{{\varPhi }_{3}}(r,z,s)=\frac{1}{2\pi }\int\limits_{0}^{\infty }{{{J}_{0}}(kr)\frac{\left( Q/s \right)\sinh \left( kb \right)\sinh \left[ k\left( g-z \right) \right]}{{{\varepsilon }_{1}}\cosh (kb)\sinh (kg)+{{\varepsilon }_{2}}\sinh (kb)\cosh (kg)}}\text{d}k
\eeq
\noindent

The resistive layer (in the middle) can be shrunk and be considered infinitely thin with a constant – given surface resistivity $R_\mathrm{s}$. Therefore, the obtained configuration becomes a 2-layer geometry where ${{\varepsilon }_{1}}={{\varepsilon }_{0}}{{\varepsilon }_{r}}$ and ${{\varepsilon }_{2}}={{\varepsilon }_{0}}+1/{{R}_{s}}{{z}_{2}}s$. 
In the limit ${{z}_{2}}\to 0$ the second term becomes infinite. It can be shown that, for $z_2\rightarrow 0$ the expressions for calculating the electric potential tend also to infinity. To overpass this difficulty, a smart mathematical trick is used in \cite{riegler} based on the expression of $\varepsilon_2$  

\beq
{{\varepsilon }_{2}}={{\varepsilon }_{0}}+\frac{1}{R_s z_2 s}=\varepsilon_0+\frac{k}{\left(kz_2\right)R_s s}
\eeq
\noindent

In this form we observe that the term $k/\left(kz_2\right)R_s s$ will be finite if only if $kz_2=1$. 
In the subsequent integration by using the Bessel function, of first kind - zero order,
an accurate result is achieved for an upper limit of integration chosen at least as multiples of exponent coefficients,
as in the case of $1/z_2$, where the upper limit of the integration $k\rightarrow \infty$ corresponds to $z_2\rightarrow 0$.
However, on one hand, by this mathematical trick a infinitely thin resistive layer is considered which is surely unrealistic
experimentally. On the other hand, the calculations become remarkably simpler and informative regarding the impact of the associated parameters. Therefore, replacing $kz_2=1$, and thus $\varepsilon_0+k/R_s s$, the functions in the integral in $s$-domain are written

\beq
{{A}_{3}}(k,s){{e}^{kz}}+{{B}_{3}}(k,s){{e}^{-kz}}=\frac{Q}{{{\varepsilon }_{0}}}\frac{\sinh \left[ k\left( g-z \right) \right]}{\left[ {{\varepsilon }_{r}}\coth (kb)\sinh (kg)+\cosh (kg) \right]\left( s+1/\tau (k) \right)}
\eeq
\noindent

and in time domain

\beq
{\mathcal{A_3}}(k,t){{e}^{kz}}+{\mathcal{B_3}}(k,t){{e}^{-kz}}=\frac{Q}{{{\varepsilon }_{0}}}\frac{\sinh \left[ k\left( g-z \right) \right]{{e}^{-t/\tau (k)}}}{\left[ {{\varepsilon }_{r}}\coth (kb)\sinh (kg)+\cosh (kg) \right]}
\eeq
\noindent
where $\mathcal{A_3}(k,t)$ and $\mathcal{B_3}(k,t)$ are the inverse Laplace Transformations of $A_3(k,s)$ and $B_3(k,s)$ functions, respectively. The decay time defined above, $\tau (k)$, is given by

\beq
\tau (k)=\frac{{{R}_{s}}{{\varepsilon }_{0}}\left[ {{\varepsilon }_{r}}\coth (kb)\sinh (kg)+\cosh (kg) \right]}{k\cosh (kg)}=\frac{{{R}_{s}}{{\varepsilon }_{0}}\left[ {{\varepsilon }_{r}}\coth (kb)\tanh (kg)+1 \right]}{k}
\eeq
\label{eq_decay}
\noindent

The obtained expression of the potential in time domain becomes

\beq
{{\varPhi }_{3}}(r,z,t)=\frac{Q}{2\pi {{\varepsilon }_{0}}}\int\limits_{0}^{\infty }{{{J}_{0}}(kr)\frac{\sinh \left[ k\left( g-z \right) \right]{{e}^{-t/\tau (k)}}}{{{\varepsilon }_{r}}\coth (kb)\sinh (kg)+\cosh (kg)}}\text{d}k
\eeq
\noindent

\begin{figure}[!ht]
\centering
\includegraphics[width=12.0cm]{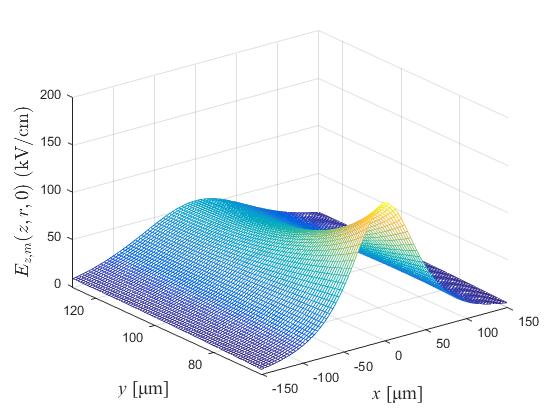} 
\caption[]{The electric field $E_\mathrm{z,m}(z)$ as a function of $z$ (which is y in the plot) due to the presence of the Micromesh calculated within a small range around $x\approx 0$. The lower value of $z$ has been set to $64\,\upmu$m which is in the middle of the amplification region.}
\label{ez_mesh}
\end{figure}

\begin{figure}[!ht]
\centering
\begin{minipage}[b]{0.47\linewidth}
\centering
\includegraphics[scale=0.40]{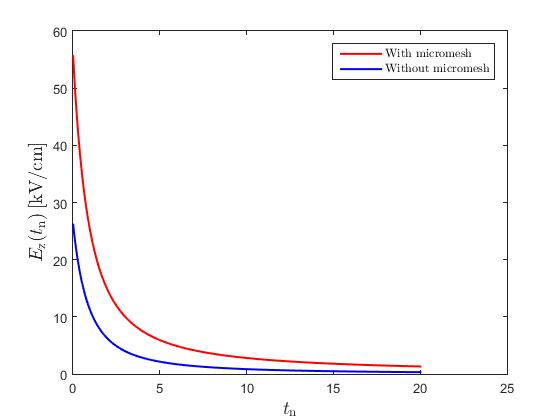}  
\caption[]{Plot of $E_\mathrm{z}$ with (red curve) and without Micromesh (blue curve) bas a function of normalized time $t_\mathrm{n}$ at $z=128\, \upmu\mathrm{m}$ and $r=0$.}
\label{ez_decay}
\end{minipage}
\qquad
\begin{minipage}[b]{0.47\linewidth}
\centering
\includegraphics[scale=0.40]{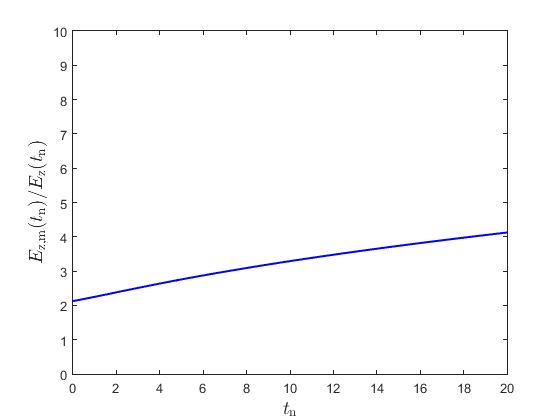}
\caption[]{The ratio of $E_\mathrm{z}$ with and without Micromesh as a function of normalized time $t_\mathrm{n}$ at $z=128\, \upmu\mathrm{m}$ and $r=0$.}
\label{ez_comp}
\end{minipage}
\end{figure}

And the component $E_{3,\text{z}}$ should be

\beq
{{E}_{3,\text{z}}}(r,z,t)=-\frac{\partial {{\varPhi }_{3}}(r,z,t)}{\partial z}=\frac{Q}{2\pi {{\varepsilon }_{0}}}\int\limits_{0}^{\infty }{k{{J}_{0}}(kr)\frac{\cosh \left[ k\left( g-z \right) \right]{{e}^{-t/\tau (k)}}}{{{\varepsilon }_{r}}\coth (kb)\sinh (kg)+\cosh (kg)}}\text{d}k
\eeq
\noindent
The electric field $E_\mathrm{z,m}(z)$ as a function of $z$ (using y in the plot) due to the presence of the Micromesh calculated within a small range around $x\approx 0$ is shown in Fig. \ref{ez_mesh}. The above expression can be also simplified by using the dimensionless variable $\kappa=kb$, as follows

\beq
{{E}_{3,\text{z}}}(r,z,t)=\frac{Q}{2\pi {{\varepsilon }_{0}}{{b}^{2}}}\int\limits_{0}^{\infty }{\kappa {{J}_{0}}(\kappa r/b)\frac{\cosh \left[ \kappa \left( g-z \right)/b \right]{{e}^{-t/\tau (k)}}}{{{\varepsilon }_{r}}\coth (\kappa )\sinh (\kappa g/b)+\cosh (\kappa g/b)}}\text{d}\kappa
\eeq
\noindent
where the corresponding expression of decay time becomes

\beq
\tau (\kappa /b)=\frac{{{R}_{s}}{{\varepsilon }_{0}}b\left[ {{\varepsilon }_{r}}\coth (\kappa )\tanh (\kappa g/b)+1 \right]}{\kappa}
\eeq
\noindent

The electric field $E_\mathrm{z,m}$ has been calculated by using numerical integration, while regarding the decay time, given in Eq. \ref{eq_decay}, we used the approximation given in Appendix \ref{appendix}, $\tau(\kappa) \approx T'/e^{-\kappa(1-e^{-2\kappa})}$, and assuming that $g/b>1$ and in turn $\tanh(\kappa g/b)\approx 1$ mostly for higher values of $\kappa$ where the resulted integral is accurate. 
The obtained ratio of $E_\mathrm{z,m}$ (with Micromesh) with respect to $E_\mathrm{z,c}$ (without Micromesh) is shown as a function of $x$ in Fig. \ref{ez_comp} at the more distant observation point above the deposited point charge, at Micromesh plane ($z=128\,\upmu$m). The actual electric field, due to grid-shape of the Micromesh, is suspected to be in between the above values and unity, depending on the position of the observation point, if it is close to the metallic wires or around the openings of the Micromesh.

The presence of the grounded Micromesh provides much higher electric field while the decay times are $t_\mathrm{n}=1.3$ and $t_\mathrm{n}=1.2$, respectively. These results are shown in Fig. \ref{ez_comp} and Fig. \ref{ez_decay}, calculated at $r=0$.
By the above approximation, the expression of decay time $\tau(\kappa)$ takes the same form as in the case of absence of Micromesh. and the decay times are very close together.

\subsection{Calculations by using 3-layer geometry}
\label{s5-2}
The calculations performed in the above Sections have been based on the approximation based on 2-layer geometry, according to \cite{riegler}, where the resistive layer is assumed of infinitely small thickness. By using this methodology the calculations are
simpler, however, the consideration of infinitely thin layer, $z_2\equiv d \rightarrow 0$ (we use the symbol $d$ in the next) is
useful for understanding the electric field creation mechanism, but it is unrealistic because a) the deposition of such a thin 
layer is impossible even if the required material exists, and b) the decay time of the obtained transient electric field 
will be constantly very small, and in turn, not adjustable. For this reason we were forced to apply the accurate methodology 
based on 3-layer geometry including the resistive layer (film) in the middle and giving to it the actual thickness value which
is $d=15\,\mathrm{\upmu}$m. Moreover, these calculations helped us to verify that the two methodologies are precisely consistent
at values of thickness $d$ down to $1$\,nm.

\makeatletter
\newcommand{\vast}{\bBigg@{4}}
\newcommand{\Vast}{\bBigg@{4}}
\makeatother


\subsection{Comparisons with the 2-layer geometry approximation}
For applying the 3-layer geometry it is necessary to formulate the 6-equation linear system which becomes 5-equation
linear system in the case without including the Micromesh. We start finding out the corresponding determinant
setting the appropriate quantities for the point charge, deposited on the resistive layer at $t=0$, and the
permittivity of this layer, as follows

\begin{eqnarray}
Q_1 &=& Q/s \nonumber\\
Q_2 &=& 0 \nonumber\\
\varepsilon_2(s) &=& \varepsilon_0 + 1/(s R_s d) \nonumber
\end{eqnarray}
\noindent
where $R_s$ is the surface resistivity and $s$ the complex variable in the Laplace domain.
Assuming in this stage, $\varepsilon_1=\varepsilon=\varepsilon_0$ we formulate the 5 equations of the linear system
as follows: 

According to the boundary condition of the potential at $z=-b$ we have
\begin{eqnarray}
{{\left. {{\varPhi }_{1}} \right|}_{z=-b}}=0\Rightarrow {{A}_{1}}{{e}^{-kb}}+{{B}_{1}}{{e}^{kb}}=0
\end{eqnarray}
\noindent

Also, the boundary conditions of the potential at $z=0$ and $z=d$ (taking also $A_3=0$, because of finite potential at $z\rightarrow \infty$), we write

\begin{eqnarray}
{{\left. {{\varPhi }_{1}} \right|}_{z=0}}={{\left. {{\varPhi }_{2}} \right|}_{z=0}}\Rightarrow {{A}_{1}}+{{B}_{1}}={{A}_{2}}+{{B}_{2}}
\label{e1}
\end{eqnarray}
\noindent

\begin{eqnarray}
{{\left. {{\varPhi }_{2}} \right|}_{z=d}}={{\left. {{\varPhi }_{3}} \right|}_{z=d}}\Rightarrow {{A}_{2}}{{e}^{kd}}+{{B}_{2}}{{e}^{-kd}}=0{{e}^{kd}}+{{B}_{3}}{{e}^{-kb}}={{B}_{3}}{{e}^{-kb}}
\label{e2}
\end{eqnarray}
\noindent

Additionally, from the Gauss law at $z=0$ we can write

\begin{eqnarray}
{{\varepsilon }_{1}}\left( {{A}_{1}}-{{B}_{1}} \right)-{{\varepsilon }_{2}}\left( {{A}_{2}}-{{B}_{2}} \right)={{Q}_{1}=Q/s}
\label{e3}
\end{eqnarray}
\noindent

And also at $z=d$

\begin{eqnarray}
{{\varepsilon }_{2}}\left( {{A}_{2}}{{e}^{kd}}-{{B}_{2}}{{e}^{-kd}} \right)-{{\varepsilon }_{0}}\left( 0{{e}^{kd}}-{{B}_{3}}{{e}^{-kd}} \right)={{Q}_{2}}=0
\label{e4}
\end{eqnarray}
\noindent

Combining the above equations Eq.~\eqref{e1}, \eqref{e2}, \eqref{e3}, \eqref{e4} and rearranging the corresponding terms, we obtain the following determinant

\begin{align}
D(k,s)=
\begin{vmatrix}
 \exp (-k b) & \exp (k b) & 0 & 0 & 0 \\
 1 & 1 & -1 & -1 & 0 \\
 {\varepsilon_0} & -{\varepsilon_0} & -{\varepsilon_2} & {\varepsilon_2} & 0 \\
 0 & 0 & \exp (k d) & \exp (-k d) & -\exp (-k d) \\
 0 & 0 & {\varepsilon_2} \exp (k d) & -{\varepsilon_2} \exp (-k d) & {\varepsilon_0} 
\exp (-k d) \\
\end{vmatrix}
\end{align}

Concentrating in the calculation of the potential of layer 3, we need only the expression of the constant $B_3(k,s)$.
We have

\begin{align}
B_3(k,s)\times D(k,s)=
\begin{vmatrix}
 \exp (-k b) & \exp (k b) & 0 & 0 & 0 \\
 1 & 1 & -1 & -1 & 0 \\
 {\varepsilon_0} & -{\varepsilon_0} & -{\varepsilon_2} & {\varepsilon_2} & Q_1 \\
 0 & 0 & \exp (k d) & \exp (-k d) & 0 \\
 0 & 0 & {\varepsilon_2} \exp (k d) & -{\varepsilon_2} \exp (-k d) & Q_2 \\
\end{vmatrix}
\end{align}

And performing the inverse Laplace Transformation

\begin{equation}
B_3(k,t)={\cal{L}}^{-1}\left[B_3(k,s)\right]\notag
\end{equation}

and by using the Mathematica framework we obtain the analytical function given below

\newgeometry{
 left=10mm,
 top=100mm}

\begin{align}
B_3(k,t) &= Q \frac{e^{2 b k}-1}{4 {\varepsilon_0} \sqrt{d^2 {\varepsilon_0}^2 {R_s}^2 
\left(4 e^{2 k (2 b+d)}-2 e^{2 d k}+e^{4 d k}+1\right)}}\times\notag\\      
&\exp \left(-\frac{t e^{-2 k (b+d)} \left(\sqrt{d^2 {\varepsilon_0}^2 {R_s}^2 
\left(4 e^{2 k (2 b+d)}-2 e^{2 d k}+e^{4 d k}+1\right)}-d {\varepsilon_0} {R_s} 
\left(-2 e^{2 k (b+d)}+e^{2 d k}-1\right)\right)}{4 d^2 {\varepsilon_0}^2 {R_s}^2}-2 
b k\right)\times\notag\\
&\Vast[d {\varepsilon_0} {R_s} \left(2 e^{2 k (b+d)}+e^{2 d k}-1\right) \left(\exp 
\left(\frac{t e^{-2 k (b+d)} \sqrt{d^2 {\varepsilon_0}^2 {R_s}^2 \left(4 e^{2 k (2 
b+d)}-2 e^{2 d k}+e^{4 d k}+1\right)}}{2 d^2 {\varepsilon_0}^2 
{R_s}^2}\right)-1\right)+\notag\\
&\sqrt{d^2 {\varepsilon_0}^2 {R_s}^2 \left(4 e^{2 k (2 b+d)}-2 e^{2 d k}+e^{4 d 
k}+1\right)} \left(\exp \left(\frac{t e^{-2 k (b+d)} \sqrt{d^2 {\varepsilon_0}^2 
{R_s}^2 \left(4 e^{2 k (2 b+d)}-2 e^{2 d k}+e^{4 d k}+1\right)}}{2 d^2 
{\varepsilon_0}^2 {R_s}^2}\right)+1\right)\Vast]\notag
\end{align}

\restoregeometry

\begin{figure}[!ht]
\centering
\begin{minipage}[b]{0.47\linewidth}
\centering
\includegraphics[scale=0.41]{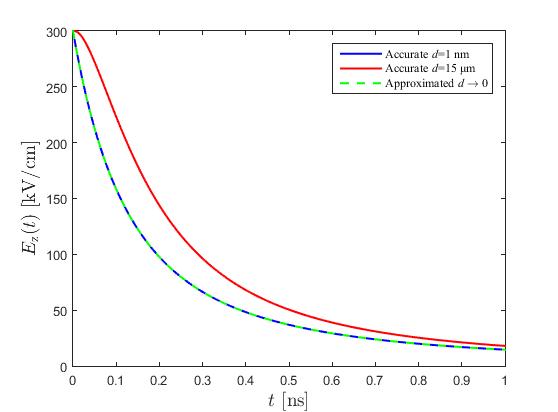}  
\caption{Plot of the electric field $E_\mathrm{z}$ versus time, at $r=10\,\upmu \mathrm{m}$ and $z=64\,\upmu \mathrm{m}$, for comparing among three cases of calculations: the accurate of 3-layer geometry with $d=1$\,nm (blue), the accurate of 3-layer geometry with $d=15\,\upmu \mathrm{m}$ (red) and the approximation of 2-layer geometry with $d\rightarrow 0$ (dashed green).}
\label{p6}
\end{minipage}
\qquad
\begin{minipage}[b]{0.47\linewidth}
\centering
\includegraphics[scale=0.41]{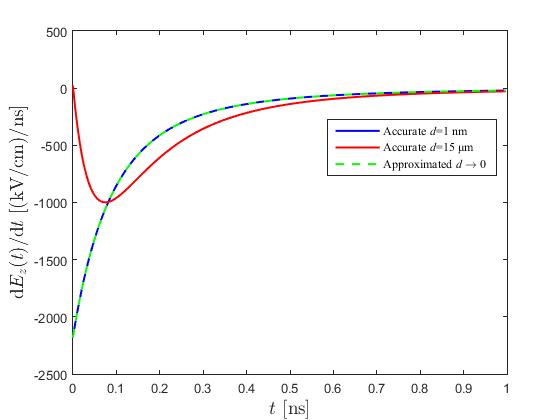}
\caption{Plot of the derivative $\mathrm{d}E_\mathrm{z}(t)/\mathrm{d}t$ versus time, at $r=10\,\upmu \mathrm{m}$  and $z=64\,\upmu \mathrm{m}$, for comparing among three cases:  the accurate of 3-layer geometry with $d=1$\,nm (blue), the accurate of of 3-layer geometry with $d=15\,\upmu \mathrm{m}$ (red), the approximation of 2-layer geometry with $d\rightarrow 0$ (dashed green).}
\label{p66}
\end{minipage}
\end{figure}

In Fig.\ref{p6} the electric field $E_\mathrm{z}(t)$ versus time is shown for comparison in three cases. In Fig. \ref{p66} the
corresponding derivatives are also plotted. The derivatives express the decay rate in different moments during the field drop. 
The blue and dashed curves, regarding the magnitude as well as the decay rate, coincide verifying that the approximation method is very accurate at the range $d\leq1$\,nm. The red curve shows a its minimum (maximum decay rate) at about $0.75$\,ns, but at larger times the decay rate is slightly smaller than the other two curves, tending to be equalized after $1$\,nm.
The electric field at $0.75$\,ns is $36\%$ higher with $d=15\,\upmu \mathrm{m}$ than that with $d=1\, \mathrm{nm}$. This is the systematic relative error of the 2-layer geometry approximation when the resistive layer thickness is finite and at the scale of the Micromegas detectors.


\begin{figure}[!ht]
\centering
\begin{minipage}[b]{0.47\linewidth}
\centering
\includegraphics[scale=0.41]{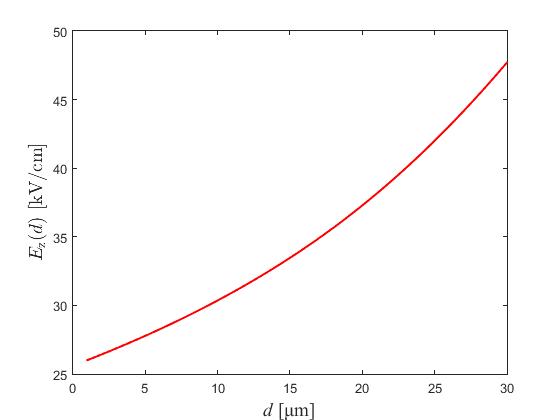}  
\caption{Plot showing the dependence of the electric field versus the resistive layer thickness $d$ within a realistic range at $r=0$ and $t=1\,$ns, assuming a given surface resistivity (constant) $R_\mathrm{s}$. The smaller the thickness the smaller slope of the curve showing asymptotic behaviour towards to $1\,$nm and beyond it.}
\label{p1}
\end{minipage}
\qquad
\begin{minipage}[b]{0.47\linewidth}
\centering
\includegraphics[scale=0.41]{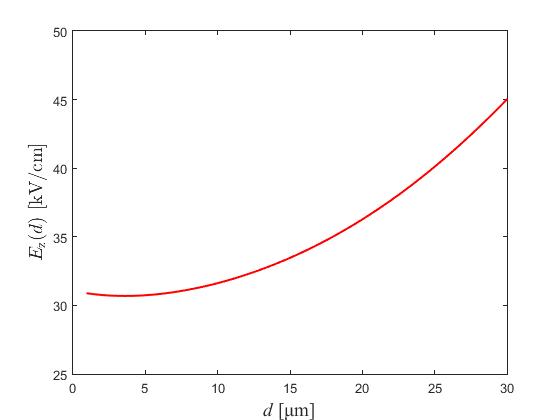}
\caption{Plot showing the dependence of the electric field versus the resistive layer thickness $d$ within a realistic range at $r=0$ and $t=1\,$ns, assuming a given resistivity (constant) $\rho=3.6\,\mathrm{\Omega\,m}$. The minimum of the electric field at $d=3.6\,\upmu\mathrm{m}$ can help for counterbalancing the uncertainty of the thickness during deposition.}
\label{p22}
\end{minipage}
\end{figure}

\subsection{Field mapping results by 3-layer geometry}
After all, and understanding the charge spreading mechanism versus time and the effect of the resistive layer thickness we present the calculations regarding the electric field including all the parameters of the NSW Micromegas detectors. Namely, including the grounded Micromesh at $z=128\,\upmu \mathrm{m}$, the actual value of relative permittivity of the insulator layer ($\varepsilon_r=3.55$) and the actual resistive layer thickness ($d=15\,\upmu \mathrm{m}$). 

These calculations are based on formulating of the 6-equation linear system. The corresponding determinant is given below.

\begin{align}
D(k,s)=
\begin{vmatrix}
 \exp (-k b) & \exp (k b) & 0 & 0 & 0 & 0 \\
 1 & 1 & -1 & -1 & 0 & 0 \\
 {\varepsilon_1} & -{\varepsilon_1} & -{\varepsilon_2} & {\varepsilon_2} & 0 
& 0 \\
 0 & 0 & \exp (k d) & \exp (-k d) & -\exp (k d) & -\exp (-k d) \\
 0 & 0 & {\varepsilon_2} \exp (k d) & -{\varepsilon_2} \exp (-k d) & 
-{\varepsilon_0} \exp (k d) & {\varepsilon_0} \exp (-k d) \\
 0 & 0 & 0 & 0 & \exp (k g) & \exp (-k g) \\
\end{vmatrix}
\end{align}

The electric field is calculated according to the integral that we have already used in other Sections, but the constants $A_3(k,s)$ and $B_3(k,s)$ have been calculated numerically followed by performing the inverse Laplace Transformation using, again, Mathematica framework. Having the calculation procedure completed, we thought it worthwhile to study first the dependence on the electric field from the thickness within a range from $d=1\,\upmu \mathrm{m}$ to $d=30\,\upmu \mathrm{m}$. 

The first study has been performed assuming a given (constant) surface resistivity ($R_\mathrm{s}=0.24\,\mathrm{M\Omega}$). The second has been performed assuming a given (constant) resistivity $\rho=3.6\,\mathrm{\Omega\,m}$. The resulting plots are shown in Figs. \ref{p1} and \ref{p22}, respectively. From view point of production procedure of a resistivity layer, the first case concerns the choice of the resistive material in order to satisfy the condition $\rho=R_\mathrm{s}d$ for the desired thickness. The second case concerns the choice of the thickness and afterwards calculating the resulting surface resistivity $R_\mathrm{s}=\rho/d$ for experimental confirmation.

\beq
{{E}_{3,\text{z}}}(r,z,t)=-\frac{\partial {{\Phi }_{3}}(r,z,t)}{\partial z}=\frac{Q}{2\pi {{\varepsilon }_{0}}}\int\limits_{0}^{\infty }{k{{J}_{0}}(kr)\left( {{A}_{3}}{{e}^{kz}}+{{B}_{3}}{{e}^{-kz}} \right)}\text{d}k
\noindent
\label{e5}
\eeq

\begin{figure}[!ht]
\centering
\begin{minipage}[b]{0.47\linewidth}
\centering
\includegraphics[scale=0.41]{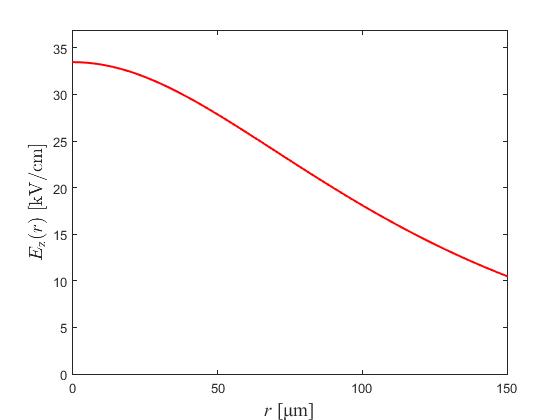}  
\caption{Plot of the electric field $E_\mathrm{z}$ versus radial distance $r$ at $z=64\,\upmu \mathrm{m}$ and $t=1\,\mathrm{ns}$, calculated by 3-layer geometry with $d=15\,\upmu \mathrm{m}$.}
\label{p2}
\end{minipage}
\qquad
\begin{minipage}[b]{0.47\linewidth}
\centering
\includegraphics[scale=0.41]{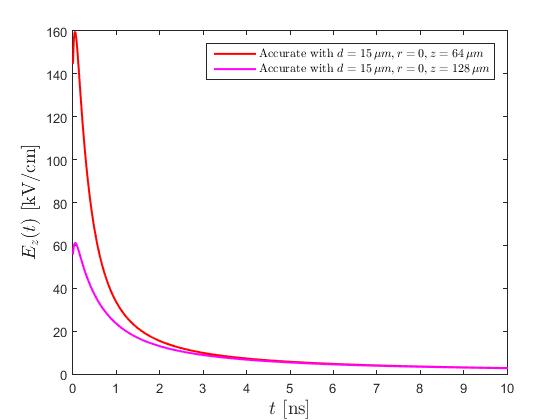}
\caption{Plot of the electric field $E_\mathrm{z}$ versus time at $r=0$ calculated by 3-layer geometry with $d=15\,\upmu \mathrm{m}$.}
\label{p3-4}
\end{minipage}
\end{figure}

In Figs. \ref{p2} and \ref{p3-4}, the electric field versus radial and vertical distance are shown, respectively.

\section*{Conclusions}
In the present work, analytic studies of the spreading mechanism of the surface electric point charge deposited on the resistive strips of the ATLAS NSW Micromegas detectors have been performed. Based on the relevant literature we went further introducing a useful approximation for the charge density for small values of time, in the range of a few ns. Additionally, the transmission line approach has been compared, by means of its equivalence with that of surface one, mainly for large values of time greater than $44$\,ns.

The methodology is for studying the component $E_\mathrm{z}$ of the created transient electric field and its mapping versus time, can be applied in different geometries, 2-layer and 3-layer, which is the accurate one. By studying the effect of the resistive layer thickness in its realistic scale, we found that it affects significantly the decay rate of the electric field. 
We have also shown that, due to the charge spread on the resistive strip, the overall electric field in the amplification region present a transient radial valley, roughly equal to the width of the resistive strip even within a time slice of 1.1\,ns. In the case of smaller deposited point charge, this valley will be shallower, accordingly. Moreover, the presence of the Micromesh increases the transient electric field $E_\mathrm{z}$ at every time, mostly above $z=64\,\upmu \mathrm{m}$. Regarding the impact of the insulator width $b$: the greater the width (e.g. $+50\%$) the higher electric field ($+14\%$). 

\section*{Author Contributions} 
The authors have contributed equally.

\vskip 0.2cm

\section*{Funding} 
This work was funded in part by the Basic Research Program PEVE 2021 of the National Technical University of Athens and the U.S. Department of Energy, Office of Science, High Energy Physics under Contracts DE-SC0012704.

\section*{Data availability statement}
This manuscript has no associated data or the data will not be deposited.

\section*{Declarations}

\section*{Conflict of interest}
The author has declared he has no conflict of interest with regard to this content.

\section*{Ethics}
The author has declared ethics committee/IRB approval is not relevant to this content.

\noindent

\section*{Acknowledgements}
We also gratefully thank Emmanuel Dris, Professor Emeritus, for his valuable comments and discussions.

\appendix\newpage\markboth{Appendix}{{Appendix}}
\renewcommand{\thesection}{\Alph{section}}
\section{Electric Field without Micromesh}
\label{appendix}

In this Appendix we present the analytical calculations based on 2-layer geometry approximation regarding the configuration of a pair of resistive and readout strip in NSW Micromegas detector without including Micromesh. The analysis is based on the methodology given in \cite{riegler}. The result given in this work concerns the charge density versus time in a different configuration not including an insulator between the resistive layer and the ground plane. Unlikely, in our case, an insulator having permittivity $\varepsilon=\varepsilon_r\varepsilon_0$ the electric potential is used. Let us consider the configuration given in Fig. \ref{figapp1}. 
The starting point is to find the potential in the region $z>0$, with the constant $A_3(,k,t)=0$, which in cylindrical coordinates is written

\beq
{{\varPhi }_{3}}(r,z,t)=\frac{1}{2\pi }\int\limits_{0}^{\infty }{{{J}_{0}}(kr)\left[ {{B}_{3}}(k,t){{e}^{-kz}} \right]}\text{dk}
\label{equapp1}
\eeq
\noindent

The expression for the constant $B_{3}$, assuming $g\rightarrow \infty$, as below  

\beq
{{B}_{3}}=\underset{g\to \infty }{\mathop{\lim }}\,\frac{2\left( Q/s \right)\sinh \left( kb \right)}{4\left[ {{\varepsilon }}\cosh (kb)\sinh (kg)+{{\varepsilon }_{2}}\sinh (kb)\cosh (kg) \right]}{{e}^{kg}}=\frac{\left( Q/s \right)\sinh (kb)}{{{\varepsilon }}\cosh (kb)+{{\varepsilon }_{2}}\sinh (kb)}
\label{equapp2}
\eeq
\noindent

Because the thin resistive layer is infinitely thin with a given surface resistivity, we must set $\varepsilon_2=\varepsilon_0+1/sR_sz_2$, where $z_2$ is the width of the very thin resistive layer. This width must be infinitely small, that is $z_2\rightarrow 0$. By using the mathematical trick $1/z_2=k$, the upper integration limit (of infinity) gives an accurate result satisfying the condition  $z_2\rightarrow 0$. Replacing $\varepsilon_2$ we obtain

\beq
{{B}_{3}}=\frac{\left( Q/s \right)\sinh (kb)}{{{\varepsilon }}\cosh (kb)+\left[ {{\varepsilon }_{0}}+\left(k/R_{s}s\right)\right]\sinh (kb)}=\frac{Q{{R}_{s}}\sinh (kb)}{{{R}_{s}}{{\varepsilon }_{0}}s\left[ {{\varepsilon }_{r}}\cosh (kb)+\sinh (kb) \right]+k\sinh (kb)}
\label{equapp3}
\eeq
\noindent

Performing the inverse Laplace transform we can write the previous expression as

\beq
{{B}_{3}}=\frac{Q{{R}_{s}}\sinh (kb)}{{{R}_{s}}{{\varepsilon }_{0}}s\left[ {{\varepsilon }_{r}}\cosh (kb)+\sinh (kb) \right]+k\sinh (kb)}=\frac{Q}{{{\varepsilon }_{0}}}\frac{1}{\left[ {{\varepsilon }_{r}}\coth (kb)+1 \right]\left( s+1/\tau (k) \right)}
\label{equapp4}
\eeq
\noindent
where

\beq
\tau (k)=\frac{{{R}_{s}}{{\varepsilon }_{0}}\left[ {{\varepsilon }_{r}}\cosh (kb)+\sinh (kb) \right]}{k\sinh (kb)}=\frac{{{R}_{s}}{{\varepsilon }_{0}}\left[ {{\varepsilon }_{r}}\frac{\cosh (kb)}{\sinh (kb)}+1 \right]}{k}=\frac{{{R}_{s}}{{\varepsilon }_{0}}\left[ {{\varepsilon }_{r}}\coth (kb)+1 \right]}{k}
\label{equapp5}
\eeq
\noindent

is a time constant in time the domain which can be written more analytically

\beq
\tau (k)=\frac{{{R}_{s}}{{\varepsilon }_{0}}\left[ {{\varepsilon }_{r}}\frac{{{e}^{2bk}}+1}{{{e}^{2bk}}-1}+1 \right]}{k}=\frac{{{\varepsilon }_{r}}+1}{2\left( 1-{{e}^{-2bk}} \right)k\nu}+\frac{\left( {{\varepsilon }_{r}}-1 \right){{e}^{-2bk}}}{2\left( 1-{{e}^{-2bk}} \right)k\nu}
\label{equapp6}
\eeq
\noindent
where $\nu=1/2\varepsilon_0R_s$.

The time constant can be written in such a way to be referred with that of the $\varepsilon=\varepsilon_0$

\beq
{{\left. \tau (k) \right|}_{{{\varepsilon }}={{\varepsilon }_{r}}{{\varepsilon }_{0}}}}={{\left. \tau (k) \right|}_{{{\varepsilon }}={{\varepsilon }_{0}}}}\left[ \frac{{{\varepsilon }_{r}}+1}{2}+\frac{\left( {{\varepsilon }_{r}}-1 \right){{e}^{-2bk}}}{2} \right]={{\left. \tau (k) \right|}_{{{\varepsilon }}={{\varepsilon }_{0}}}}\frac{1}{2}\left[ {{\varepsilon }_{r}}+1+\left( {{\varepsilon }_{r}}-1 \right){{e}^{-2bk}} \right]
\label{equapp7}
\eeq
\noindent
We observe that it depends on $k$, that is, ${{\left. \tau (k\to 0) \right|}_{{{\varepsilon }}={{\varepsilon }_{r}}{{\varepsilon }_{0}}}}={{\varepsilon }_{r}}{{\left. \tau (k) \right|}_{{{\varepsilon }}={{\varepsilon }_{0}}}}$ and
${{\left. \tau (k\to \infty ) \right|}_{{{\varepsilon }}={{\varepsilon }_{r}}{{\varepsilon }_{0}}}}=\frac{{{\varepsilon }_{r}}+1}{2}{{\left. \tau (k) \right|}_{{{\varepsilon }}={{\varepsilon }_{0}}}}$.

The second value is the smaller and thus corresponds to the worst case in our study. For this reason we proceed with this expression which is also a good approximation for $k>1$. Coming again to the function $B_3$ we have

\begin{eqnarray}
{{B}_{3}}(k,t)=\frac{Q}{{{\varepsilon }_{0}}}\frac{1}{{{\varepsilon }_{r}}\coth (kb)+1}{{e}^{-t/\left\{ {{\left. \tau (k) \right|}_{{{\varepsilon }} 
{{\varepsilon }_{0}}}}\left[ {{\varepsilon }_{r}}+1+\left( {{\varepsilon }_{r}}-1 \right){{e}^{-2bk}} \right] \right\}}}
\label{equapp7a}
\end{eqnarray}
\noindent
leading to

\begin{eqnarray}
{{B}_{3}}(k,t)\approx \frac{Q}{{{\varepsilon }_{0}}}\frac{1}{{{\varepsilon }_{r}}\coth (kb)+1}{{e}^{-k\left( 1-{{e}^{-2bk}} \right)2\nu t/(\varepsilon_r+1)}}
\label{equapp8}
\end{eqnarray}
\noindent

It can be also written, using the dimensionless variable $\kappa=kb$, as follows 

\begin{eqnarray}
{{B}_{3}}(\kappa ,t)\approx\frac{Q}{{{\varepsilon }_{0}}}\frac{1}{{{\varepsilon }_{r}}\coth (\kappa )+1}{{e}^{-\kappa \left( 1-{{e}^{-2\kappa }} \right)t/{{T}'}}}
\label{equapp9}
\end{eqnarray}
\noindent
where ${T}'=\frac{{{\varepsilon }_{r}}+1}{2}T=\frac{{{\varepsilon }_{r}}+1}{2}2b{{\varepsilon }_{0}}{{R}_{s}}=b\left( {{\varepsilon }_{r}}+1 \right){{\varepsilon }_{0}}{{R}_{s}}=b\left( {{\varepsilon }_{0}}+{{\varepsilon }} \right){{R}_{s}}$
\noindent
Moreover, we can perform the following approximation for small times $t_\mathrm{n}=t/T'<<20$, we can approximate $e^{-2kb}<<1$, leading to ${{\varepsilon }_{r}}\cot (bk)+1={{\varepsilon }_{r}}\left(1+{{e}^{-2bk}}\right)/\left(1-{{e}^{-2bk}}\right)+1\approx {{\varepsilon }_{r}}+1$.

\noindent
Therefore, the final expression for $B_3$ is

\beq
{{B}_{3}}(\kappa ,t)\approx\frac{Q}{{{\varepsilon }_{0}}\left( {{\varepsilon }_{r}}+1 \right)}{{e}^{-\kappa \left( 1-{{e}^{-2\kappa }} \right)t/T'}}=\frac{Q}{\left( {{\varepsilon }_{0}}+{{\varepsilon }} \right)}{{e}^{-\kappa \left( 1-{{e}^{-2\kappa }} \right)t/T'}}
\label{equapp10}
\eeq
\noindent

The potential is written

\beq
{{\varPhi }_{3}}(r,z,t)=\frac{1}{2\pi b^2}\int\limits_{0}^{\infty }{{{J}_{0}}\left( \kappa r/b \right)\left[ {{B}_{3}}(\kappa /b,t){{e}^{-\kappa z/b}} \right]}\text{d}\kappa
\label{equapp11}
\eeq
\\
\noindent
The other two constants, $A_1$ and $B_1$ (we notice that $A_3=0$), are calculated according to the same methodology and are given below

\beq
{{A}_{1}}(\kappa ,z,t)\approx \frac{Q}{{{\varepsilon }_{0}}+{{\varepsilon }}}\left( 1-{{e}^{-2\kappa b}} \right){{e}^{-\kappa \left( 1-{{e}^{2\kappa }} \right)t/T'}}
\label{equapp12}
\eeq
\\
\noindent 

\beq
{{B}_{1}}(k,z,t)=-\frac{Q}{{{\varepsilon }_{0}}+{{\varepsilon }}}{{e}^{-2kb}}{{e}^{-k\left( 1-{{e}^{-2kb}} \right)t/T'}}
\label{equapp13}
\eeq
\\
\noindent 
The charge density can be calculated according to the Gauss law

\beq
{{\left. q(r,t)={{\varepsilon }_{0}}{{\varepsilon }_{r}}\frac{\partial {{\varPhi }_{1}}(r,z,t)}{\partial z} \right|}_{z=0}}{{\left. -{{\varepsilon }_{0}}\frac{\partial {{\varPhi }_{3}}(r,z,t)}{\partial z} \right|}_{z=0}}={{\varepsilon }_{0}}\left[ {{\left. {{\varepsilon }_{r}}\frac{\partial {{\varPhi }_{1}}(r,z,t)}{\partial z} \right|}_{z=0}}{{\left. -\frac{\partial {{\varPhi }_{3}}(r,z,t)}{\partial z} \right|}_{z=0}} \right]
\label{equapp14}
\eeq
\\
\noindent
and after replacing the constants takes the form

\beq
q(r,t)=\frac{{Q{\varepsilon }_{0}}}{2\pi }\int\limits_{0}^{\infty }{k{{J}_{0}}(kr)\left[ {{\varepsilon }_{r}}{{A}_{1}}+\left( -{{\varepsilon }_{r}}{{B}_{1}}+{{B}_{3}} \right) \right]}\text{d}k
\label{equapp15}
\eeq
\\
\noindent
We also calculate

\beq
-{{B}_{1}}+{{B}_{3}}=\frac{Q{{\varepsilon }_{0}}}{{{\varepsilon }_{0}}+\varepsilon }\left\{ \left( {{\varepsilon }_{r}}-1 \right)\left[ {{e}^{-2kb}}{{e}^{-k\left( 1-{{e}^{-2kb}} \right)t/T'}} \right]+{{e}^{-k\left( 1-{{e}^{-2kb}} \right)t/T'}} \right\}
\label{equapp16}
\eeq
\\
\noindent
leading to 

\begin{eqnarray}
-{{B}_{1}}+{{B}_{3}}\approx \frac{Q{{\varepsilon }_{0}}}{{{\varepsilon }_{0}}+{{\varepsilon}}}{{e}^{-k\left(1-{{e}^{-2kb}}\right)t/T'}}
\label{equapp17}
\end{eqnarray}
\\
\noindent
after performing the approximations, $e^{-2kb}<<1$ for $k>>1/2b$. Therefore, the charge density becomes

\begin{eqnarray}
q(r,t)\approx \frac{Q{{\varepsilon }_{0}}}{2\pi {{b}^{2}}\left( {{\varepsilon }_{0}}+{{\varepsilon }} \right)}\int\limits_{0}^{\infty }{\kappa {{J}_{0}}(\kappa r/b)\left[ {{\varepsilon }_{r}}{{e}^{-\kappa \left( 1-{{e}^{-2\kappa }} \right)t/T'}}+{{e}^{-k\left( 1-{{e}^{-2\kappa }} \right)t/T'}} \right]}\text{d}\kappa
\label{equapp18}
\end{eqnarray}
\\
\noindent
which is also written

\begin{eqnarray}
q(r,t)\approx \frac{Q{{\varepsilon }_{0}}\left( \varepsilon_r+1 \right)}{2\pi {{b}^{2}}\left( {{\varepsilon }_{0}}+{{\varepsilon }} \right)}\int\limits_{0}^{\infty }{\kappa {{J}_{0}}(\kappa r/b){{e}^{-k\left( 1-{{e}^{-2\kappa }} \right)t/T'}}}\text{d}\kappa
\label{equapp18a}
\end{eqnarray}
\\
\noindent

And finally

\beq
q(r,t)\approx \frac{Q}{2\pi {{b}^{2}}}\int\limits_{0}^{\infty }{\kappa {{J}_{0}}(\kappa r/b){{e}^{-\kappa \left( 1-{{e}^{-2\kappa }} \right)t/T'}}}\text{d}\kappa
\label{equapp19}
\eeq
\\
\noindent

The electric field is obtained by using the potential, as follows

\beq
{{E}_{3,\text{z}}}(r,z,t)=-\frac{\partial {{\varPhi }_{3}}(r,z,t)}{\partial z}=\frac{Q}{2\pi {{b}^{2}}{{\varepsilon }_{0}}}\int\limits_{0}^{\infty }{\kappa {{J}_{0}}(\kappa r/b)\frac{{{e}^{-\kappa z/b}}{{e}^{-\kappa \left( 1-{{e}^{-2\kappa }} \right)t/T'}}}{{{\varepsilon }_{r}}\coth (\kappa )+1}}\text{d}\kappa 
\label{equapp20}
\eeq
\\

\end{document}